\documentclass[a4paper,11pt]{article}
\pdfoutput=1 

\usepackage{jcappub} 
\usepackage[T1]{fontenc}
\graphicspath{{./}{figures/}}
\usepackage[title]{appendix}

\title{Hubble distancing: Focusing on distance measurements in cosmology}

\author{Kylar L. Greene}
\author{and Francis-Yan Cyr-Racine}
\affiliation{Department of Physics and Astronomy, University of New Mexico \\
210 Yale NE,
Albuquerque, NM 87131, USA}

\emailAdd{kygreene@unm.edu}
\emailAdd{fycr@unm.edu}

\abstract{
The Hubble-Lema\^{i}tre tension is currently one of the most important questions in cosmology. Most of the focus so far has been on reconciling the Hubble constant value inferred from detailed cosmic microwave background measurement with that from the local distance ladder. This emphasis on one number -- namely $H_0$ -- misses the fact that the tension fundamentally arises from disagreements of distance measurements. To be successful, a proposed cosmological model must accurately fit these distances rather than simply infer a given value of $H_0$. 
Using the newly developed likelihood package `\textit{distanceladder}', which integrates the local distance ladder into \texttt{MontePython}, we show that focusing on $H_0$ at the expense of distances can lead to the spurious detection of new physics in models which change late-time cosmology. As such, we encourage the observational cosmology community to make their actual distance measurements broadly available to model builders instead of simply quoting their derived Hubble constant values.}

\begin{document}
\maketitle
\flushbottom

\section{Introduction} \label{sec:intro}

The pioneering work of Slipher, Lema\^{i}tre, Robertson, Leavitt, and Hubble in the late 1920s established that the Universe is expanding \cite{Leavitt:1912zz,Slipher1915,Lemaitre:1927zz,Robertson1928,Hubble:1929ig}. 
Since then, cosmological distance measurements have been at the forefront of understanding the expansion history of the Universe. 
Measuring astronomical distances is not a trivial task. 
Every cosmological distance measurement relies on establishing an absolute \textit{dimensionful} length scale that other distance measurements are then compared to.
Fortunately, we live in a Universe where multiple such absolute distances are available: photon-baryon sound horizon \cite{Evslin:2017qdn,Aylor:2018drw,Jedamzik:2020zmd,Pogosian:2020ded,Bernal:2020vbb}, parallax distances to nearby stars \cite{Semeniuk:2000na,Gaia:2016zol,Bailerjones2015}, distances to eclipsing binaries \cite{Paczynski:1996dj,Bonanos:2006jd,Graczyk:2003zy}, distances to water masers \cite{Reid:2019tiq,pesce2020,Dzib2018}, and time-delay distances to strong gravitational lenses \cite{Blandford:1991xc,Eulaers:2011aa,Arendse:2019itb}. 

However, there exists a tension between the various absolute distance scales used to infer the Hubble constant $H_0$ within the standard $\Lambda$ cold dark matter ($\Lambda$CDM) model \cite{Bernal:2016gxb,Verde:2019ivm,Knox:2019rjx,Efstathiou:2020wxn,Efstathiou:2021ocp,Paul:2021abc}. Most notably, cosmic microwave background (CMB) measurements \cite{Planck:2018vyg,aiola:2020,SPT-3G:2021eoc} and the Cepheid-calibrated local distance ladder \cite{Riess:2011yx,freedman2012,Riess:2016jrr,CSP:2018rag,Riess:2020fzl} infer Hubble constant values that are discrepant at the critical $5\sigma$ level \cite{Riess:2021jrx}. On the other hand, an alternative calibration of the local distance ladder based on the tip of the giant branch (TRGB) \cite{Beaton:2016nsw,Hatt:2017rxl,Hatt:2018opj,Hatt:2018zfv,Hoyt2019vi,Beaton2019vii,jang2021ix,Hoyt_2021,Freedman:2019jwv,Freedman:2020dne,Freedman:2021ahq} leads to a Hubble constant value that is somewhat intermediate between that inferred from the CMB and the Cepheid-calibrated ladder, while strong-lensing time delays \cite{Treu:2016ljm,Suyu:2016qxx,Birrer:2018vtm,Wong:2019kwg} generally find $H_0$ values that are consistent with the Cepheid-calibrated distance ladder, albeit with larger possible systematic errors \cite{Birrer:2020tax}. 
A major question in cosmology now is whether this discrepancy is the result of a yet-to-be-discovered systematics or is caused by a breakdown of our current cosmological paradigm \cite{Krishnan:2021dyb,Krishnan:2021jmh,Luongo:2021nqh}.

Reference \cite{DiValentino:2021izs} summarizes many possible new-physics solutions to alleviate the tension, while ref.~\cite{Schoneberg:2021qvd} ranks the proposed models with respect to a common data set. Generally speaking, new-physics scenarios can be classified into two broad categories, depending on whether they primarily modify the early or late Universe. 
Early-time solutions aim to decrease the value of the sound horizon $r_{\rm{s}}$ \cite{Evslin:2017qdn,Aylor:2018drw,Jedamzik:2020zmd,Pogosian:2020ded,Bernal:2020vbb}, either by injecting energy into the pre-recombination Universe (see e.g.~refs.~\cite{Karwal:2016vyq,Poulin:2018cxd,Smith:2019ihp,Smith:2020rxx,Poulin:2021bjr,Agrawal:2019lmo,Cyr-Racine:2021alc,Murgia:2020ryi,Krishnan:2020obg,Blinov:2020hmc,Blinov:2020uvz,Niedermann:2020dwg,Choi:2020pyy,Kreisch:2019yzn,Brinckmann:2020bcn,Das:2020xke,Mazumdar:2020ibx,Aloni:2021eaq,Sakstein:2019fmf,CarrilloGonzalez:2020oac}) or by other means \cite{Jedamzik:2020abc,Rashkovetskyi:2021rwg,Sekiguchi:2020teg,Hart:2020} in such a way that also preserves the CMB measurements. Late-time solutions aim to increase the value of $H_0$ locally by modifying the expansion history at $z \lesssim 1$ (see e.g.~refs.~\cite{Mortonson:2009qq,Dhawan:2020xmp,Jassal:2005qc,Caldwell:2003vq,DiValentino:2020naf,Mortonson:2009qq,Benevento:2020fev,Alestas:2020zol,Keeley:2019esp,Dutta:2018vmq,Alestas:2021luu,Clark:2020miy,Braglia:2020iik,Ballardini:2020iws,Karwal:2021vpk,Nunes:2021zzi,Yang:2018euj,Yang:2018qmz,Yang:2018uae,Yang:2020ope,Yang:2020uga,Yang:2021flj,Yang:2021eud,Yang:2021hxg,DiValentino:2016hlg,DiValentino:2017rcr,DiValentino:2019exe,DiValentino:2019ffd,DiValentino:2019jae,DiValentino:2020kha,DiValentino:2020kpf,DiValentino:2020leo,DiValentino:2020vnx,DiValentino:2021rjj,DiValentino:2021zxy}).
At face value, these late solutions may appear to resolve the tension because a sudden increase in the Hubble expansion rate at very-late times could yield a larger apparent $H_0$ value while only causing a small fractional change in the overall distance to the CMB last-scattering surface.
However, such solutions are misleading in the notion that the local distance ladder \emph{directly} measures the Hubble expansion rate at $z = 0$ \cite{Efstathiou:2021ocp}.
In practice, the local distance ladder infers $H_0$ from measurements of the peak absolute type Ia supernova (SNe Ia) magnitude $M_{\rm{sn}}$ and of Hubble flow SNe Ia with redshift $z \gtrsim 0.02$. To be successful, late-time solutions therefore need to accurately fit the measured distances to Hubble-flow supernovae, rather than trying to obtain a given $H_0$ value derived using $\Lambda$CDM assumptions. 

Indeed, the singular focus on $H_0$ rather than on the calibrated distances to Hubble flow objects often obscures what model ingredients are necessary to address the root cause of the discrepancy \cite{Lin:2019htv,Lin:2021sfs,Garcia-Quintero:2019cgt,Pan:2019hac,Camarena:2021jlr,Knox:2019rjx}. 
In cosmological models that are phenomenologically close to $\Lambda$CDM at late times (including models that only affect the early Universe), this focus on $H_0$ is warranted as distances to Hubble-flow objects are inversely proportional to the Hubble constant, and the cosmographic expansion of the luminosity distance is accurate at low redshifts. 
For these, using a Gaussian prior on $H_0$ to represent the entire distance ladder is likely sufficient, albeit at the price of discarding information about the actual goodness of fit of model distances to low-redshift objects. 
On the other hand, for cosmologies that differ significantly from $\Lambda$CDM at late times, the relationship between $H_0$ and distances is more complex as other model parameters could enter the computation of the latter. 
In this case, simply focusing on $H_0$ can be extremely misleading as it is not possible to boil down the entirety of the distance ladder to a single number without important loss of information. 
Instead, the distances to Hubble-flow objects must be directly accounted for in assessing the success of such cosmological scenarios.   

In this paper, we show that the Hubble-Lema\^{i}tre tension is not simply about a single number -- namely $H_0$ -- but rather about a large array of distance measurements at low redshifts. 
In this language, ``solving'' this discrepancy is really about finding a self-consistent expansion history of the Universe that can describe all cosmological data, including the CMB, large-scale structure, Big-Bang Nucleosynthesis abundance yields, baryon acoustic oscillation, and the distances to nearby Hubble-flow objects such as SNe Ia and strong lenses.
Using the newly developed likelihood package \textit{distanceladder}\footnote{The package is publicly available at \href{https://github.com/kylargreene/distanceladder}{distanceladder}.}, which provides a fast yet accurate fit to the entirety of the local distance ladder, we show that apparently promising late-time solutions that have been presented in the literature, in fact, provide poor fits to the actual measured distances to low-redshift objects.
Even more concerning, we show that some evidence for new physics that has been reported in the literature using a Gaussian $H_0$ prior appears to be spurious once the actual distance measurements are properly taken into account. 
Our work strongly indicates that cosmological analyses should focus on fitting cosmological distances rather than argue about the ``correct'' value of the current expansion rate. 

This paper is structured as follows. 
Section \ref{sec:ldlreview} discusses the various methods in which distances are measured locally and the absolute dimensionful scale they establish. 
It also reviews how the local distance ladder is built. Section \ref{sec:dllikelihood} presents the details of the \textit{distanceladder} package we have developed to correctly analyze changes to late-time cosmology and also present the various consistency checks that we have performed. 
In section \ref{sec:LDE}, we use this likelihood package to demonstrate that a simple Gaussian $H_0$ prior cannot capture the complexity of the entire local distance ladder and can even lead to spurious detection of new physics. 
We finally conclude in section \ref{conclusion}.

\section{Distance measurements in cosmology} \label{sec:ldlreview}
In cosmology, any distance measurement depends on first establishing an absolute dimensionful anchor that defines the problem's overall scale.
Such anchors do not necessarily need to be absolute distance scales, but can also be a time interval, the linear size of an object, an acceleration, or a temperature, among others. What they all have in common is that they establish a dimensionful scale from which a distance can eventually be measured. In the following, we briefly review different distance measurement techniques in cosmology, highlighting the absolute \textit{dimensionful scale} anchoring each approach, and describe the logical flow from that anchor to the inferred cosmological distances. Our goal here is to provide a concise end-to-end description of the distance measurements that are instrumental to the existence of the Hubble-Lema\^{i}tre Tension. As such, we focus below on the local distance ladder, the cosmic microwave background, and strong gravitational lensing cosmography. Other important techniques include cosmic chronometers \cite{Jimenez:2001gg,Stern:2009ep,Crawford:2010rg,Melia:2013hsa,Wei:2016ygr,Ruan:2019icc,Borghi:2021rft}, gravitational wave standard sirens \cite{Holz:2005df,Nishizawa:2012vk,LIGOScientific:2016aoc,LIGOScientific:2017adf,Belgacem:2018lbp}, surface brightness fluctuations \cite{Blakeslee:1998pb,Liu:2001nc,Khetan:2020hmh,Blakeslee:2012fi,Fritz:2012mu,Jensen:1998bi}, Tully-Fisher relation \cite{Giovanelli:1996zv,Sakai:1999aw,Tutui:2001vc,Bonhomme:2008ti,Sofue:1996nu,Russell:2008it,Kourkchi_2020}, Type II supernovae \cite{Gall:2017gva,Blinnikov:2012xb,deJaeger:2020zpb,Hamuy:2003tc,SNLS:2006mwe,deJaeger:2016cev}, and HII galaxies \cite{Chavez:2012km,Leaf:2017dcx,Yennapureddy:2017vvb,Wei:2016jqa}. 

\subsection{The Distance Ladder} \label{sec:distanceladder}

\subsubsection{Anchor Measurements} \label{sub:anchor}

In cosmology, distance ladders are built by establishing the absolute distance to nearby stars or galaxies. 
Such distances usually referred to as anchors, are derived from measurements of other absolute dimensionful scales. 
Common methods of measuring the distance to these anchors include stellar parallax \cite{Casertano:2015dso,vanLeeuwen:2007xw,Riess:2018uxu,Madore:1997va,jones2018,Perryman:1997sa,fabricius2021}, detached eclipsing binary stars \cite{Pietrzynski:2013gia,Vilardell:2009aa,southworth2015debcat,Semeniuk:2001bd,Paczynski:1996dj}, and MASER emission from the accretion disk of super-massive black holes \cite{Reid:2019tiq,wiggins2016,Tarchi:2012kg,Pihlstrom:2004vg,pesce2020,Gao:2015tqd,Kuo:2012hg}. We briefly describe these below, noting the key dimensionful scales entering the problem in each case.

\paragraph{Parallax} The modern parallax method measures the apparent movement of a stellar object with respect to the background stars on the sky as the Earth revolves around the sun.
Because the star is very far away compared to the perceived motion of the star on the night sky, the parallax equation is simply
\begin{equation}
    d = \frac{1}{p},
\end{equation}
where $d$ is the distance in parsecs, and $p$ is the parallax angle measured in arcseconds.
The absolute \textit{dimensionful scale} for the modern parallax method is the baseline distance established between two observations, typically the diameter of Earth's orbit.
This is a precisely measured value known as an astronomical unit (AU) equivalent to $149,597,870,700\pm3$ meters \cite{Pitjeva2009ProposalsFT,luzum2011}.
While parallax measurements are highly precise, they are severely limited in the measurable distance, with the furthest measurements reaching the Milky Way's galactic centre using the Gaia space telescope \cite{Gaia:2016zol,vanLeeuwen2017,luri2018,torra2021}.

\paragraph{Detached Eclipsing Binaries} The Detached Eclipsing Binary (DEB) method utilizes binary stars whose orbit takes the pair within the observer's line of sight, resulting in primary and secondary eclipses \cite{Paczynski:1996dj,pietrzynski2019distance,Southworth:2005xi,Bonanos:2006jd,southworth2015debcat,remple2021determining}. 
Here, the pair's ``detached'' nature means that the binary stars' separation is much larger than their individual radii, which avoids complications related to mass transfer and accretion disks. 
The absolute \textit{dimensionful scale} entering the problem is the orbital period of the binary pair, which can be measured both from the light curve and from radial velocity spectroscopic data. 
These observations can also be used to measure the inclination and eccentricity of the DEB system's orbit, which, when combined with the orbital period measurement, allows one to determine the orbital radius from Kepler's laws. 
Then, the shape of the primary and secondary eclipses can be used to calculate the radius of each star.
Finally, spectroscopic and color information is used to estimate the surface brightness of each star, which can be used to compute the distance to the pair
\begin{equation}\label{eq:dist_DEB}
    d = \left(\frac{F_1}{F_{\rm{1,tel}}}\right)^{1/2}R_1 = \left(\frac{F_2}{F_{2,\rm{tel}}}\right)^{1/2}R_2
\end{equation}
where $F_i$ is the flux from each star, $F_{i,\rm{tel}}$ is the flux received at the telescope, and $R_i$ is the radius of that star determined using the method outlined above. 
The two estimates for $d$ in Eq.~\eqref{eq:dist_DEB} provide a vital consistency check for the distance measurement. 
This procedure can also be carried in multiple photometric bands, providing further cross-checks of the distance estimate. 
The assumptions surrounding the surface brightness estimate of each star represent the most significant potential systematics of the DEB technique.
However, previous work has demonstrated consistent models for surface brightness with negligible dependence on metallicity effects \cite{Thompson:2000gi,salsi2021}.
Additionally, the systems are difficult to detect due to their detached nature, causing transits to be very short, which limits the number of available distance measurements \cite{Paczynski:1996dj}.

\paragraph{MASERS}
Super-massive black holes (SMBH) heat the surrounding gas in their accretion disk to very high temperatures, producing x-ray emission. 
This strong radiation field can stimulate various molecules present within the disk, particularly that of water, resulting in localized MASER emission \cite{reid1988,reid1988distance,Lo2005,hachisuka2006water,gwinn1992distance}.
Such MASERs are high intensity and point-like, allowing for line-of-sight (LOS) velocities to be measured in the accretion disk within thousands of Schwarzchild radii of the SMBH.
Furthermore, the orbital speed of the MASERs around the central engine is large enough that their LOS acceleration can also be measured over a few years.
Once the LOS acceleration, LOS velocity, and velocity gradient are measured, a distance can be determined according to
\begin{equation}
    d = \frac{\partial_{\theta}v_{\rm{LOS}}}{a_{\rm LOS}}v_{\rm{LOS}}
\end{equation}
where $v_{\rm{LOS}}$ is the LOS velocity, $\theta$ is the angle on the sky according to the observer, $\partial_{\theta}v_{\rm{LOS}}$ is the LOS velocity gradient on the sky, and $a_{\rm LOS}$ is the LOS acceleration \cite{haschick1994}.
In this case, the absolute \textit{dimensionful scale} established by the MASER method is the LOS acceleration.
The MASER method additionally can directly measure distances into the Hubble flow, notably by the Megamaser Cosmology Project who directly measure from distances to be $H_0 = 73.9 \pm 3.0$ km/s/Mpc \cite{pesce2020}.
However, current systematic errors associated with radio phase calibration make the MASER method challenging to perform \cite{Wijnholds:2010ks}.
Additionally, the small number of known MASERS with nearby phase calibrators limits the method's applicability.

\subsubsection{Calibrating Standard Candles}
The second step of the distance ladder depends on calibrating standard candles.
These calibrators must satisfy the following criteria; (i) have anchor measurements of their distance from us using one of the techniques outlined in section \ref{sub:anchor}, and (ii) exist in a host galaxy with other, brighter standard candles (such as SNe Ia) that are observable in the Hubble flow. 
While we focus below on cepheids and the tip of the red giant branch, we note that mira variable and RR Lyrae stars have also been used as calibrating standard candles \cite{Huang:2019yhh,Rau:2018eui,yuan2018,Muraveva2018,klein2011}.

\paragraph{Cepheid Variable Stars}  
Leavitt's law gives a relationship between the period and apparent magnitude of a Cepheid variable star \cite{Leavitt:1912zz}.
Once anchor measurements determine the distance to a subsample of Cepheids, Leavitt's Law can be calibrated.
Then, the apparent magnitude of a hypothetical Cepheid star with a one day-long period is calculated using linear regression.
The absolute magnitude of Cepheids, $M_{\rm{ceph}}$, is determined by the distance modulus equation using the anchor distance measurement of a Cepheid and the apparent magnitude of the hypothetical one day period Cepheid.
In terms of magnitude, the distance modulus $\mu$ is given by
\begin{equation} \label{eq:dmod}
    \mu = m - M
\end{equation}
where $m$ is the apparent magnitude and $M$ is the absolute magnitude.
Then, the apparent magnitude of a hypothetical one day period Cepheid in a nearby host galaxy is calculated using observations of Cepheid populations.
Finally, the distance to the host galaxy is calculated from equation \eqref{eq:dmod} using $M_{\rm{ceph}}$ and the calculated apparent magnitude of a one day period Cepheid in the same host galaxy.
Recently, the nature of Leavitt's Law has come under question as the relationship between period and magnitude for Cepheids may not be wholly linear but have a `break' where the slope changes value \cite{bhardwaj2016break,Ngeow:2005qc,Ngeow:2007qb}.
Additionally, metallicity effects on the star may change the period to luminosity relationship, although previous studies have found the overall effect from metalicity to be small \cite{Freedman:1990ab,Freedman:2011xv}.

\paragraph{Tip of the Red Giant Branch} 
Red giant branch stars are old, evolved stars with an inert helium core.
Once the inert helium core reaches conditions to allow for fusion, it ignites in a highly energetic event and expels a large amount of matter from the star's atmosphere, leading to an immediate decrease in luminosity.
This behaviour leads to the `tip of the red giant branch' (TRGB) as a standard candle, as the helium flash will occur at similar mass conditions for all red giant branch stars \cite{Salaris:1997sg}.
The peak absolute magnitude of the tip, $M_{\rm{TRGB}}$, is calculated by the CCHP (Carnegie-Chicago Hubble Program) and Freedman (2021) (hereafter referred to as F21) in a three-step process \cite{Beaton:2016nsw,Hatt:2017rxl,Hatt:2018opj,Hatt:2018zfv,Hoyt2019vi,Beaton2019vii,Freedman:2019jwv,jang2021ix,Freedman:2021ahq}.
First, a large sample of potential red giant branch stars from the same galaxy undergoes a selection cut based on colour to remove other evolved stars.
Second, the luminosity function is determined from the cut sample.
Third, a Sobel edge detection filter is used to determine the edge of the luminosity function, which corresponds to the TRGB.
Once $M_{\rm{TRGB}}$ is calculated and calibrated with an anchor measurement (using one of the techniques described in section \ref{sub:anchor}), the TRGB can be used to determine the distance to SNe Ia host galaxies much in the same way that other calibrating standard candles do.
However, only observations of the halo's of galaxies should be used to calculate $M_{\rm{TRGB}}$ as the inner regions of galaxies will have asymptotic giant branch stars which contaminate the sample \cite{Freedman:2019jwv}.

\subsubsection{SNe Ia}
The final rung of the distance ladder's involves SNe Ia.
SNe Ia occurs in binary systems in which a white dwarf is actively accreting material from its binary companion.
Eventually, the electron degeneracy pressure can no longer support the mass of the white dwarf, and the Chandrasekhar mass limit \cite{Chandrasekhar:1931ftj} is reached (approximately equal to 1.44 \(M_\odot\)), at which point the white dwarf collapses, and a supernova occurs.
These are excellent standard candles for measuring distances well into the Hubble flow
because the mass at which these supernovae occur is consistent and the event's magnitude bright.

\paragraph{Local SNe Ia} 
Measuring the absolute magnitude of SNe Ia, $M_{\rm{sn}}$, requires measurements of SNe Ia in the same host galaxy as calibrating standard candles \cite{phillips1993absolute,branch1992type}, such as cepheids or TRGB stars.
Because the distance to the host galaxy is much greater than the host galaxy's size, it is assumed that the calculated distance modulus of the calibrating standard candle is the same as the SNe Ia. Ref.~\cite{Freedman:2019jwv} describes 18 recorded SNe Ia occurring in galaxies in which TRGB measurements are available, while ref.~\cite{Riess:2021jrx} describes 42 recorded SNe Ia occurring in galaxies in which Cepheid data are available. 

\paragraph{Hubble Flow SNe Ia} The distance modulus to distant SNe Ia well into the Hubble flow is then measured using $M_{\rm{sn}}$ and apparent magnitude measurements.
These distant SNe Ia also have redshift measurements, enabling a distance modulus-redshift relationship to be defined, which in terms of luminosity distance is given by
\begin{equation}
    \mu = 5\text{log}(d_{\rm{L}}) + 25,
\end{equation}
 where $d_L$ is the luminosity distance in Mpc which is given by
\begin{equation} \label{eq:dl}
    d_{\rm{L}} = (1+z)\frac{c}{H_0}\int_0^z\frac{dz^{\prime}}{E(z^{\prime})},
\end{equation}
and $E(z) = H(z)/H_0$ is the dimensionless Hubble rate.
Therefore, we can solve eq.~\eqref{eq:dmod} for $H_0$ in terms of only observable quantities to find
\begin{equation} \label{eq:h0dist}
    \mu \propto \log_{10}\left(\frac{c}{H_0} \int_0^z\frac{dz^{\prime}}{E(z^{\prime})}\right).
\end{equation}
Equation \eqref{eq:h0dist} explicitly shows that distance measurements determine the value of $H_0$.
Indeed, calculating $H_0$ at $z = 0$ from the local distance ladder relies on distance measurements of SNe Ia in the redshift range $z\gtrsim$ 0.02, with the lower bound set to avoid the coherent flow of local SNe Ia \cite{Riess:2016jrr}.
Additionally, the overall shape of the redshift-luminosity relationship is relevant due to the redshift integral in eq.~\eqref{eq:dl} \cite{Efstathiou:2020wxn}.

\subsection{CMB Distance Measurements}
The most precise measurements of the Hubble constant have so far been obtained from detailed observations of the CMB \cite{Planck:2018vyg}. 
Light from the CMB last-scattering surface forms a two-dimensional projected map on the sky whose primary observables are temperature and polarization fluctuations on different angular scales. 
These perturbations tell us about critical physical processes in the pre-recombination era. 
By themselves, fractional temperature and polarization fluctuations seen projected on the sky do not set an absolute distance scale from which the Hubble constant can be inferred \cite{Cyr-Racine:2021alc}. 
What fundamentally allows us to determine $H_0$ from CMB observations in the $\Lambda$CDM model is the knowledge of the CMB temperature today $T_0$, which provides an \emph{absolute scale} on which cosmological distances important to the CMB can be calibrated \cite{Ivanov:2020mfr}.

The most prominent CMB angle on the sky is the angular size of the sound horizon at last scattering ($z_\star$), which is given by
\begin{equation}
    \theta_{\star} = \frac{r_{\rm{s}}}{D_{\rm{A}}(z_{\star})},
\end{equation}
where $r_{\rm{s}}$ is the comoving baryon-photon sound horizon, and $D_{\rm{A}}(z_{\star})$ is the comoving angular diameter distance to the CMB. Mathematically, $r_{\rm{s}}$ and $D_{\rm{A}}(z_{\star})$ are given by
\begin{equation}\label{eq:rs}
    r_{\rm{s}} = \int_{z_{\star}}^{\infty}\frac{c_{\rm{s}} dz}{H(z)} \quad \text{and} \quad D_{\rm{A}}(z_{\star}) = \int_0^{z_{\star}}\frac{dz}{H(z)},
\end{equation}
where $c_{\rm{s}}$ is the sound speed of the primeval plasma, and $H(z)$ is the Hubble expansion rate. On dimensional ground, the comoving sound horizon is approximately given in $\Lambda$CDM by $r_{\rm s}\simeq 10^{-4} M_{\rm pl}/T_0^2$, where $M_{\rm pl}$ is the Planck mass and the prefactor has mild dependence on the redshift of matter-radiation equality, the photon-to-baryon ratio, and $z_\star$. On the other hand, the comoving angular diameter distance scales as $D_{\rm{A}}(z_\star) \simeq (3.1 H_0)^{-1}$, which implies that the Hubble constant inferred in $\Lambda$CDM by the CMB is approximately
\begin{equation}\label{eq:H0_CMB}
    H_0 \sim 3.1\times 10^{4}\theta_{\star} \frac{ T_0^2}{ M_{\rm pl}}.
\end{equation}
While precise measurements of the anisotropies on the CMB give a measurement of $\theta_{\star}$ to an accuracy of $0.03\%$ \cite{Planck:2018vyg}, uncertainties in other (dimensionless) cosmological parameters entering the prefactor in eq.~\eqref{eq:H0_CMB} result in the CMB measuring $H_0$ to a $0.8\%$ precision within $\Lambda$CDM.

\subsection{Strong Lensing Cosmography}
Gravitational lensing cosmography offers a critical check of distance measurements, as lensing distances can directly probe distances well into the Hubble flow and are independent of the local distance ladder and the CMB\cite{Treu:2016ljm,Suyu:2016qxx,Birrer:2018vtm,Wong:2019kwg,Birrer:2020tax}.
Strong gravitational lensing occurs when a significant mass distribution is in the line of sight between the observer and the source, which results in the appearance of multiple images of the background source \cite{Bartelmann:2010fz}. If the source is variable such as a quasar, we can measure the time delay difference, $\Delta t_{\rm{AB}}$, between the light curves of the different images as the light takes different paths to the observer. This time interval forms the \emph{absolute scale} in the problem and is given by
\begin{equation}
    \Delta t_{\rm{AB}} = \frac{D_{\Delta \rm{t}}}{c}\left( \phi(\theta_{\rm{A}},\beta) -\phi(\theta_{\rm{B}},\beta) \right),
\end{equation}
where $D_{\Delta \rm{t}}$ is the time delay distance, $\phi$ are Fermat potentials, $\theta$ is the image location, and $\beta$ is the corresponding source position relative to an unperturbed path.
The time delay distance then is defined by
\begin{equation}
    D_{\Delta \rm{t}} = (1+z_{\rm{d}})\frac{D_{\rm{d}}D_{\rm{s}}}{D_{\rm{sd}}},
\end{equation}
where $z_{\rm{d}}$ is the redshift of the gravitational lens, $D_{\rm{d}}$ is the angular diameter distance to the lens, $D_{\rm{s}}$ is the angular diameter distance to the source, and $D_{\rm{sd}}$ is the angular diameter distance from the lens to the source \cite{Birrer:2019otx}.
From this, we see that
\begin{equation}
    D_{\Delta \rm{t}} \propto \frac{c}{H_0}.
\end{equation}
So far, a sample of 8 quasar lenses has been used to infer the Hubble constant using this technique. 
This sample is expected to grow significantly in the near future with the advent of wide surveys enabled by the Vera Rubin Observatory, and the Euclid satellite \cite{brough2020vera,salucci2021,Euclid:2021icp}. 
A major challenge for the method remains the accurate determination of the lensing mass models for each target and the proper modelling of line-of-sight effects \cite{Tihhonova:2017mym}. Additionally, strong lensing cosmography requires a significant amount of observation time to determine the time delay accurately \cite{Bonvin:2016crt}. 
In the past, several years of data were required to overcome microlensing variability and obtain reliable time delays, but recent studies have shown that a year of daily observations could be sufficient \cite{DES:2017uif,LSSTDarkEnergyScience:2019dgr}. 
In any case, the cadence of the Legacy Survey of Space and Time at the Rubin Observatory \cite{LSSTScience:2009jmu,Jha:2019rog,LSST:2017ags} is expected to yield a sizable sample of lensing times delays that will be used for cosmography, bringing this technique well into the systematics-dominated regime. 

\section{The \textit{distanceladder} Likelihood} \label{sec:dllikelihood}

In a significant number of cosmological analyses to date for late-time changes to cosmology (see e.g.~refs.~\cite{Zhao:2017cud,Yang:2021flj,DiValentino:2020naf,Guo:2018ans,Dai:2020rfo,DiValentino:2017rcr,Yang:2020ope,DiValentino:2016hlg,DiValentino:2020vnx,Benaoum:2020qsi}), all of the richness and complexity of the distance ladder measurements described in the previous section are reduced to a simple Gaussian constraint on $H_0$ (see Refs.~\cite{Camarena:2021jlr,Alestas:2020zol} for exceptions to this). 
However, such Gaussian prior on the $z=0$ Hubble rate is an obvious over-simplification since it neglects that all of the information on the expansion rate gathered from the distance ladder comes from redshifts higher than zero. In particular, all information on the shape of the SNe Ia magnitude-redshift relation is lost when boiling down the whole distance ladder to a simple Gaussian prior on $H_0$ \cite{Verde:2016ccp,Efstathiou:2021ocp}. It is thus worth emphasizing that to resolve the Hubble-Lema\^{i}tre Tension; it is in general \emph{not} sufficient for a cosmological model to have a $z=0$ Hubble rate compatible with the value reported by the distance ladder measurements. Instead, the model must be able to successfully fit the \emph{calibrated distances} to SNe Ia in the Hubble flow. Indeed, while there is a one-to-one relation between distances and the Hubble constant for  $\Lambda$CDM models with fixed acceleration and jerk parameters $q_0$ and $j_0$ \cite{Dhawan:2020xmp}, this is in general not the case for models with a sudden late-time transition for which the cosmographic expansion of the luminosity distance breaks down. 
Such theories can appear to resolve the tension by having a Hubble constant compatible with that inferred from the distance ladder but do not necessarily provide a good fit to the actual supernova distances. 

To remedy this, we introduce here the \textit{distanceladder} numerical likelihood package, which replaces the standard Gaussian prior on $H_0$ (or $M_{\rm{sn}}$) used in many analyses with a direct fit to the calibrated distances of the SNe Ia in the Hubble flow. Our goal here is not to replace the detailed analyses performed in, e.g.~, refs.~\cite{Riess:2011yx,Riess:2016jrr,Freedman:2019jwv,Riess:2021jrx}, but instead provide the user with an end-to-end distance-ladder package such that the impact of different analysis assumptions and choices on cosmological constraints can be fully understood. We briefly describe below the content of the numerical package and its usage and then illustrate in section \ref{sec:LDE} how a model that appears to resolve the Hubble-Lema\^{i}tre Tension can be strongly disfavored once the actual distances are taken into account, using a dynamic late-time dark energy (LDE) model as a case example. For the remainder of this discussion, when a Gaussian prior is used, it utilizes the results of R21 with $H_0$ = 73.2 $\pm$ 1.3 km/s/Mpc.

\subsection{Implementation and Usage}
As its core, the `\textit{distanceladder}' likelihood package incorporates the local distance ladder as described by section \ref{sec:distanceladder} into the Monte-Carlo-Markov-Chain tool \texttt{MontePython} \cite{Brinckmann:2018cvx} by directly comparing the measured luminosity distances of Hubble flow SNe Ia to the distances predicted by a cosmological model.
The  \textit{distanceladder} nullifies the need for a Gaussian prior on $H_0$ and allows all cosmological models, even those with baroque late-time dynamics, to properly account for the distances measured by the local distance ladder.

\paragraph{User Inputs} -- \textit{distanceladder} requires two inputs from the user; a choice of calibrating standard candles and a choice of anchor measurements.
The available options of calibrating standard candles are Cepheids, the TRGB, or a subsample in which the distances to anchor SNe Ia calculated by both methods agree to 1$\sigma$ referred to as the `concordant' sample.
The concordant option allows the user to select the SNe Ia anchors which agree on distance measurements between the TRGB and the Cepheid calibration providing a unique tool to analyze the tension seen between the two methodologies.
The choice of anchors is only available for the Cepheid calibration presently.
A future update will include the 42 Cepheid calibrators \cite{Riess:2021jrx} as well as the Pantheon+ sample for Hubble flow SNe Ia \cite{Scolnic:2021amr,Carr:2021lcj}.
Additionally, future iterations may include an option to use the Carnegie Supernova Project (CSP) Hubble flow SNe Ia data \cite{Krisciunas:2017yoe} as the Hubble flow supernova data, if there is interest from the community.

\paragraph{Calibrating Standard Candles} -- If Cepheids are chosen as the calibrating standard candle of choice, $M_{\rm{ceph}}$ is first calculated based on the choice of anchors.
Currently, the LMC and NGC4258 are supported as anchor choices.
The distance to the anchors are provided directly and are not recalculated, using the results of ref.~\cite{pietrzynski2019distance} for the LMC and ref.~\cite{Reid:2019tiq} for NGC4258.
If both anchors are chosen, $M_{\rm{ceph}}$ is calculated independently for both anchors then combined using a weighted average using data from refs.~\cite{Hoffmann:2016nvl,bhardwaj2016large}.
Once $M_{\rm{ceph}}$ is calculated, the weighted average slope of Leavitt's Law is determined using the Cepheid data directly from the 19 host galaxies presented by ref.~\cite{Riess:2016jrr} in the Wesenheit near-infrared magnitude band \cite{Madore1982}.
The correlation value $R$ between Cepheid intrinsic colour and luminosity is assumed to be $R = 0.386$ as done by S$H_0$ES.
Using the Copernican principle, the \textit{distanceladder} assumes that all Cepheids in the data sample provided will have identical slopes.
The global slope of Leavitt's Law is then applied to the Cepheid samples of the 19 SNe Ia host galaxies to determine the intercept of the relation.
As discussed previously, there may exist a break in Leavitt's law in which case this method can be generalized to include a broken power-law feature in the future.
The distance modulus given by eq.~\eqref{eq:dmod} determines the distance to the host galaxies by using $M_{\rm{ceph}}$ and the intercept of Leavitt's Law for each host galaxy.
Finally, a linear regression calculates the intercept of the relation between the distance to the host galaxies and SNe Ia's apparent magnitude.
The intercept corresponds to the absolute magnitude of SNe Ia, $M_{\rm{sn}}$.

If the TRGB is chosen as the calibrating standard candle of choice, the distance modulus to the host galaxies is not calculated by the likelihood package but instead taken from table 3 from ref.~\cite{Freedman:2019jwv}.
The TRGB method of calculating distances is more involved than the Cepheid method, so it is not explicitly included in this release but may be included in future releases if there is interest.
A linear regression using the distance modulus from table 3 of ref.~\cite{Freedman:2019jwv}, and the apparent magnitude of the local SNe Ia calculates $M_{\rm{sn}}$ in the same manner as in the Cepheid method.
In all calculations, the Hubble photometry $B$ magnitudes are used as the standard filter of choice.

If the concordant sample is chosen as the calibrating standard candle, the distance modulus again is not calculated but instead taken from table 3 of ref.~\cite{Freedman:2019jwv}.
Only distance measurements in which the TRGB and Cepheid calibration agree to less than 1$\sigma$ are considered.
This cut results in a smaller sample size of 6 local SNe Ia than the 19 local SNe Ia calibrated by Cepheids or the 18 local SNe calibrated by the TRGB.
$M_{\rm{sn}}$ is then calculated in the same manner as in the prior two methods.

\paragraph{SNe Ia and Tension Calculation} -- Regardless of calibration method, eq.~\eqref{eq:dmod} calculates the distance to Hubble flow SNe Ia using $M_{\rm{sn}}$ and apparent magnitude measurements from the Pantheon data set \cite{Scolnic:2017caz}.
Since our analysis does not rely on the cosmographic expansion of the luminosity distance, the \textit{distanceladder} uses the \textit{entire} Pantheon dataset with no upper bound on redshift to ensure the shape of the redshift-luminosity relationship is fully accounted for. Separately from the \textit{distanceladder} likelihood, the Boltzmann code CLASS \cite{Lesgourgues:2011re} then calculates the luminosity distances to these Hubble flow SNe Ia for the cosmological model under consideration. 
The Hubble-Lema\^{i}tre Tension (if present in the model) is then quantified as a disagreement of distances, not of $H_0$ values, by comparing the distance moduli of the Hubble flow SNe Ia calculated from $M_{\rm{sn}}$ to those inferred from the cosmological model. 

\paragraph{Redshift Errors} -- Redshift errors are included whenever possible for the Hubble flow supernovae from ref.~\cite{Steinhardt:2020kul}.
Not all supernovae in Pantheon have errors that are available presently (this has been changed in the recent Pantheon+ release \cite{Scolnic:2021amr}).
The average fractional error of the 831 supernovae with errors available in the overall Pantheon data set is 0.83\%.
The average fractional error of the 184 supernovae with errors available with $z < 0.15$ (corresponding to the upper redshift cut of R21 and similar studies) in the Pantheon data set is 1.0\%.
While the fractional error is small, the redshift error does need to be incorporated in future studies.
The covariant matrix used by \textit{distanceladder} in minimizing the distance disagreement is generated by adding the observational errors associated with apparent magnitude and redshift as a diagonal matrix to a dense matrix where each element is the error associated with $M_{\rm{sn}}$, representing the systematic error.

\begin{figure}[t!] 
\includegraphics[width=12cm]{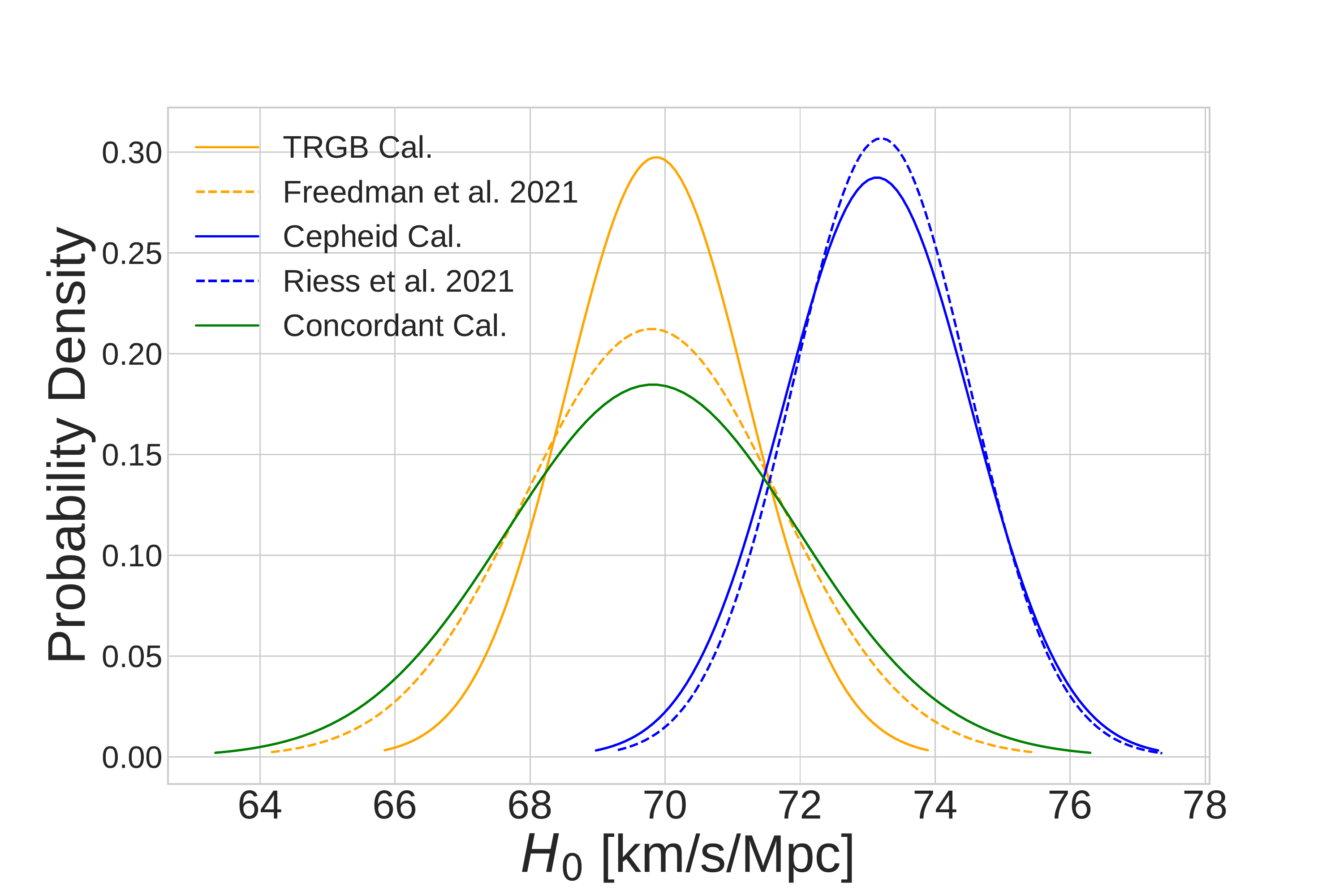}
\centering
\caption{1D posterior distributions of $H_0$ using only the \textit{distanceladder} likelihood, comparing the user input choices of using the TRGB, Cepheid, or concordant calibration. The solid lines correspond to \textit{distanceladder} results using the tip of the red giant branch calibration (yellow), Cepheid calibration (blue), or a combination of the results in which both calibration schemes agree on the host galaxy distances (green). The dashed lines correspond to the previous results obtained by ref.~\cite{Freedman:2019jwv} (yellow) and ref.~\cite{Riess:2020fzl} (blue).}
\label{fig:dlcomp}
\end{figure}

\subsection{Consistency with Previous Results} As a consistency check, figure \ref{fig:dlcomp} demonstrates the ability of the \textit{distanceladder} likelihood to recover the results of both the F21 and R21 for a $\Lambda$CDM model using the Monte-Carlo Markov Chain tool \texttt{MontePython}. 
The $\Lambda$CDM model utilizes Planck high-$\ell$ TTTEE, low-$\ell$ EE, and low-$\ell$ TT data to represent the cosmological measurement of $H_0$.
The local measurements utilize the \textit{distanceladeer} likelihood with the respective calibration choice and a prior on the baryon density to represent the local measurement of $H_0$.
The TRGB calibration accurately recovers the F21 mean value result but underestimates the error slightly as the larger Pantheon supernova data set in B magnitude is used, rather than the smaller CSP data set in the $\rm{B}^{\prime}$ magnitude, which uses a different photometric calibration scheme and different assumptions of blackbody physics.
However, because we can reproduce the mean results of the F21 study nearly precisely using the Pantheon Hubble flow supernovae with B magnitude photometry and the photometric results of ref.~\cite{Riess:2016jrr} for the anchor supernova, the results support that the CSP and Pantheon supernovae data sets are mostly consistent and do not contribute to the tension nor does the photometry of ref.~\cite{Riess:2016jrr}.
The Cepheid calibration recovers an accurate mean value of $H_0$ compared to the R21 value, with the slight difference stemming from the \textit{distanceladder} likelihood only having access to the LMC and NGC4258 as anchors while the R21 value uses the LMC, NGC4258, and Milky Way Cepheids as anchors.
Interestingly, the concordant sample finds a value of $H_0$ similar to the F21 result with larger error bars due to less anchor SNe Ia available, which should be investigated in future studies. 

Table \ref{tab:dlresults} summarizes the values of $M_{\rm sn}$ and $H_0$ for our different calibration choices in a $\Lambda$CDM model.  
An unexpected behaviour of the results is that when increasing the error in $M_{\rm{sn}}$ as seen in the Concordant sample case, a smaller value of $H_0$ is preferred from the MCMC breaking the typical relationship between $H_0$ and $M_{\rm{sn}}$. 
As a check to determine if it was indeed the error in $M_{\rm{sn}}$ driving this difference, we reran the concordant calibration scheme artificially lowering the error in $M_{sn}$ by a factor of 2.
By forcing the error in $M_{\rm{sn}}$ to be similar to that seen in the Cepheid and TRGB calibration schemes, we find that the typical relationship between $M_{\rm{sn}}$ and $H_0$ is recovered.

\begin{table}[]
\begin{center}
 \begin{tabular}{||c | c c c ||} 
 \hline
          & Cepheid Cal.       & TRGB Cal.          & Concordant Cal.           \\ [0.5ex] 
 \hline\hline
 $M_{\rm{sn}}$ & -19.226$\pm$0.039 & -19.325$\pm$0.039 & -19.267$\pm$0.064\\ [1ex]
 \hline
 $H_0$    & 73.14$\pm$1.39  & 69.87$\pm$1.34   & 69.82$\pm$2.16  \\ [1ex]
 \hline
\end{tabular}
\end{center}
\caption{The distance ladder parameters derived from \textit{distanceladder} for the three calibration options available. $M_{\rm{sn}}$ is given in the $B$ apparent, $K$ corrected peak magnitude. $H_0$ is given in units of km/s/Mpc.
The concordant sample finds a value of $H_0$ similar to the TRGB calibration at the cost of increasing the error nearly two-fold.}
\label{tab:dlresults}
\end{table}

\begin{figure*}[t!] 
\includegraphics[width=5cm]{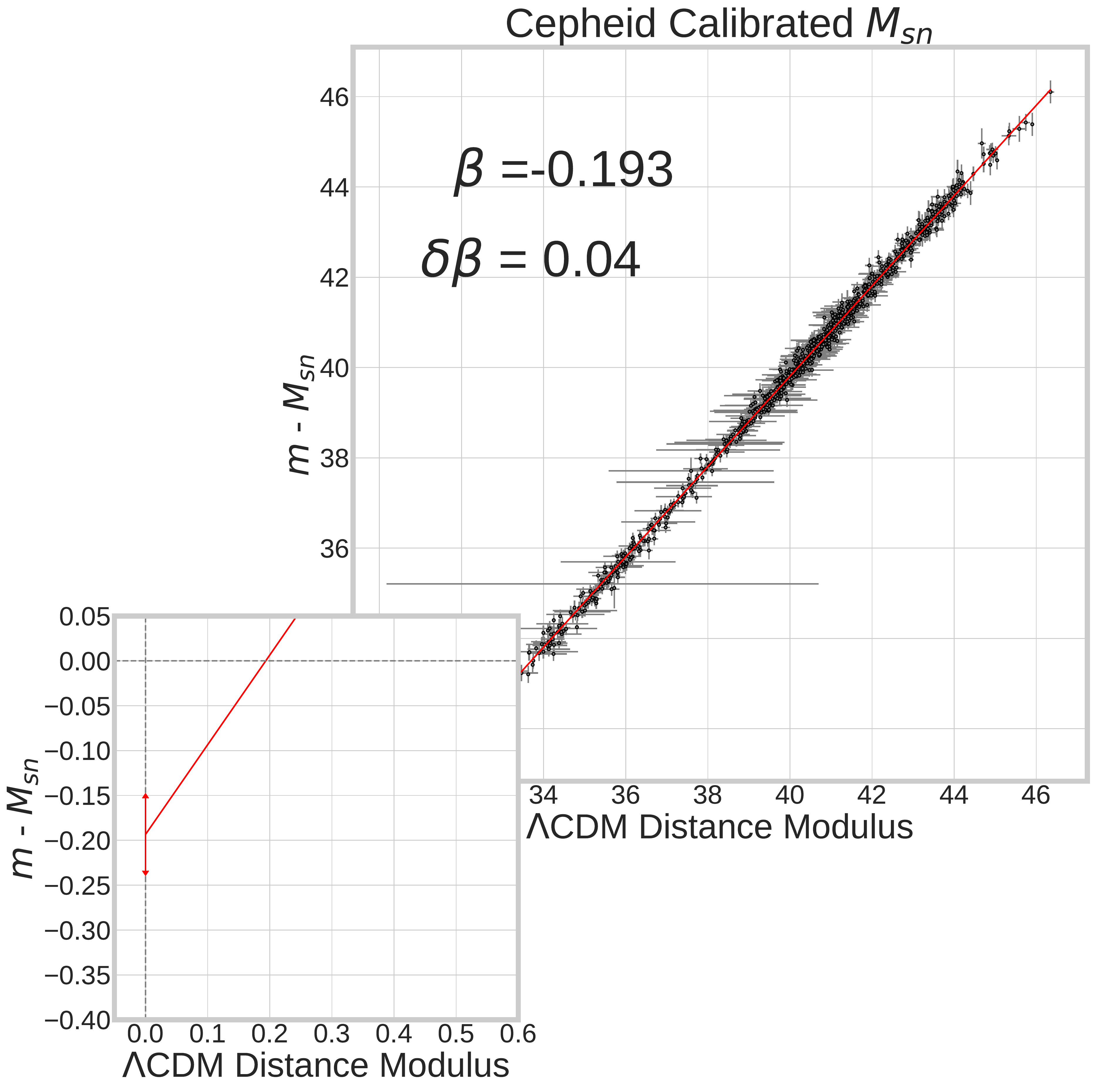}
\includegraphics[width=5cm]{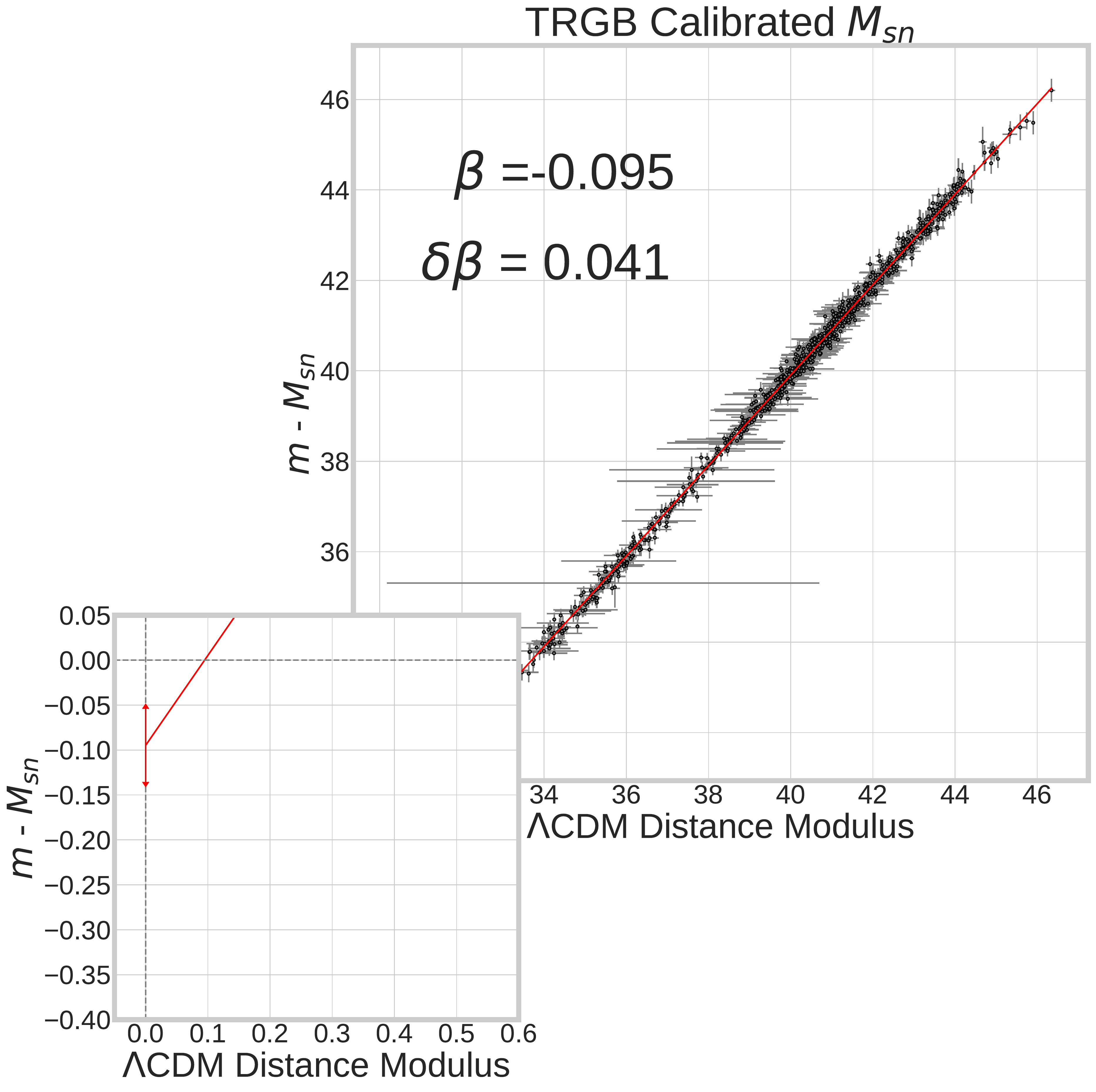}
\includegraphics[width=5cm]{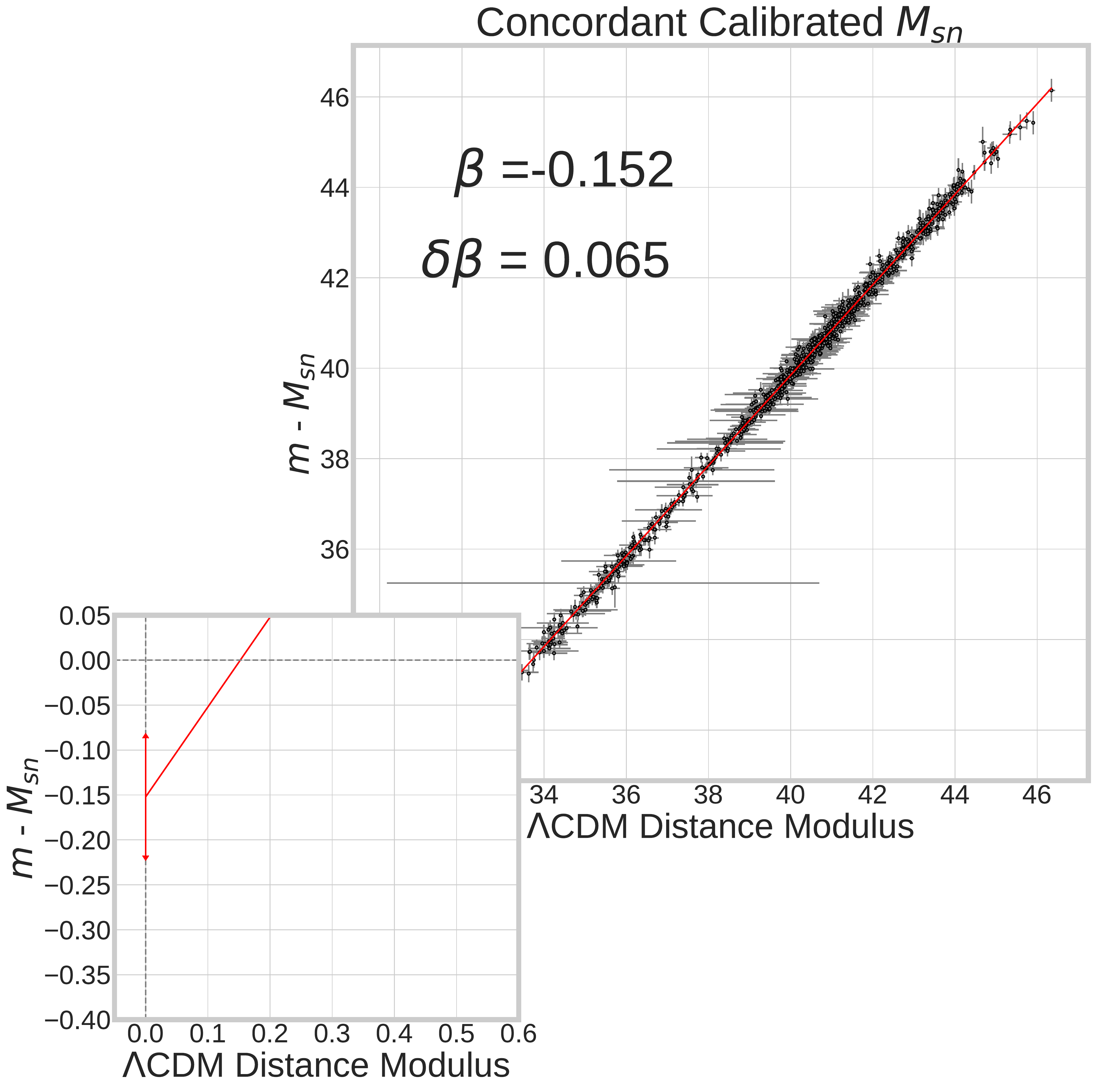}
\centering
\caption{(left) Graph comparing the calculated distance moduli for Hubble flow supernova. The $y$-axis describes the distance modulus as calculated by the local distance ladder using the $M_{\rm{sn}}$ absolute distance scale, calibrated by Cepheid variable stars. The $x$-axis describes the distance modulus as calculated from Planck CMB measurements assuming a $\Lambda$CDM model. (middle) Graph describing an identical situation as in the previous panel but using the TRGB method to calibrate $M_{\rm{sn}}$.~(right) Graph describing an identical situation as in the previous two panels, but using the concordant calibration for $M_{\rm{sn}}$.}
\label{fig:vsgraphs}
\end{figure*}

\subsection{A New Metric to assess the Tension between Data Sets} 
Now that we a have likelihood focusing on cosmological distances rather than $H_0$ itself, the next question is which metric should be used to compare local distance measurements with predictions from high-redshift observations such as the CMB? 
The intercept $\beta$ presented in figure \ref{fig:vsgraphs} may be what future studies should focus on to resolve the Hubble-Lema\^{i}tre Tension, rather than $H_0$ itself.
In this figure, the $y$-axis represents the distance moduli to Hubble flow SNe Ia as inferred by the local distance ladder.
The $x$-axis represents the distance moduli to the Hubble flow SNe Ia as inferred in a $\Lambda$CDM cosmological model fitted to Planck data, including errors associated with redshift expansion rate \cite{Planck:2018vyg}.
The intercept $\beta$ is calculated for the three calibration schemes and $\Lambda$CDM using a simple linear regression with a fixed slope of unity.
Allowing the slope to vary is possible and will be sensitive to changes in late-time cosmology, as it will modify the shape of the SN-magnitude redshift relationship, potentially detecting geometric inconsistencies.

\begin{figure}[t] 
\includegraphics[width=11cm]{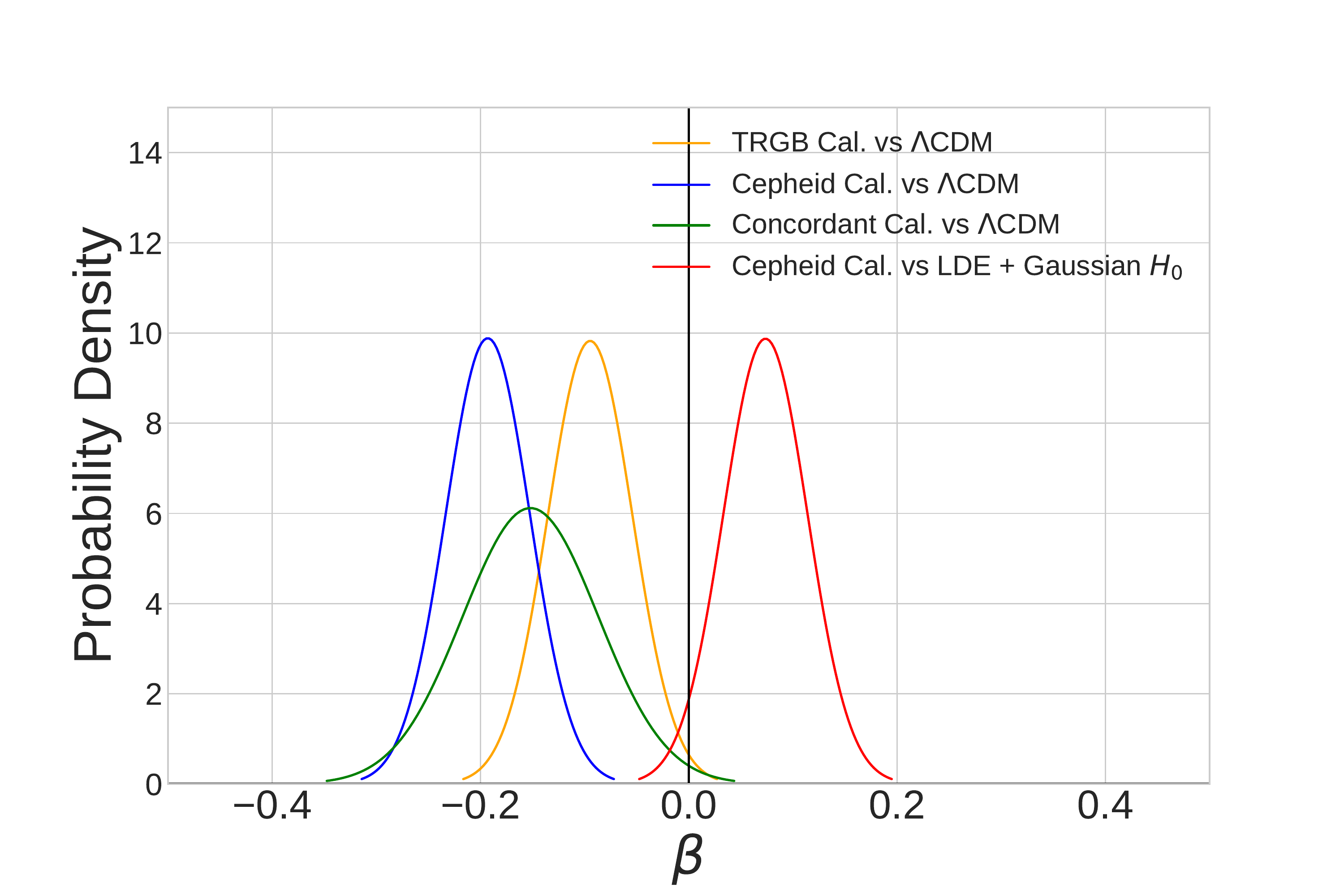}
\centering
\caption{The probability densities of $\beta$. The solid black vertical line indicates an agreement between the distance ladder and Planck CMB measurements. The blue, green, and orange lines on the left represent the tension between the chosen calibration scheme and a $\Lambda$CDM calibrated to Planck CMB data. The red line on the left represents the tension between a Cepheid calibration scheme and a late-time dark energy model using a Gaussian $H_0$ to represent the local measurements of the distance ladder. The Gaussian's are shown out to the 3$\sigma$ level.}
\label{fig:LDEvs}
\end{figure}

If no Hubble-Lema\^{i}tre Tension existed, the distance moduli established by $M_{\rm{sn}}$ and those inferred from a cosmological fit to the CMB would agree, and the intercept would be statistically consistent with zero. 
As we can see from Figure \ref{fig:LDEvs} however, the $\beta$ values for our three calibration schemes differ from zero at more than $3\sigma$ in a $\Lambda$CDM model calibrated to Planck CMB measurements, indicating the presence of the well-known Hubble-Lema\^{i}tre Tension. 
Additionally, all cases find a slope statistically identical to unity when allowing the slope to vary, indicating no geometric issues.
The width of the $\beta$ posteriors is similar as all three case examples utilize the same Pantheon data set. 
Beyond just being a parameter in linear regression, the intercept $\beta$ also has a clear physical meaning: $-\beta$ represents the change to the absolute SNe Ia magnitude necessary to bring the low-redshift distances to supernovae in agreement with those inferred from the CMB. 

\subsection{Equivalence with Gaussian Prior for Early-time Dynamics}
As an additional consistency check, we consider a simple extension of $\Lambda$CDM in which the parameter $N_{\rm{eff}}$ is allowed to freely vary. 
Since $N_{\rm{eff}}$ is a proxy for the abundance of free-streaming radiation (including neutrinos) at early times \cite{Lesgourgues:2006nd,Hannestad:2010kz}, it has only indirect impacts (i.e.~through parameter degeneracies) on the late Universe. 
As the late-time cosmology in such an extension is phenomenologically similar to $\Lambda$CDM for which the standard cosmographic expansion in terms of $q_0$ and $j_0$ is valid, we expect that the standard Gaussian $H_0$ prior and the \textit{distanceladder} likelihood should find nearly equivalent posterior distribution for the Hubble constant when fitted in combination with CMB data. 
As such, this model constitutes a simple test case to assess the equivalency of our  \textit{distanceladder} likelihood with the standard Gaussian prior on $H_0$ for cosmological models differing from $\Lambda$CDM only through their early-Universe evolution  (e.g.~ref.~\cite{Poulin:2018cxd}).

\begin{figure}[t!] 
\includegraphics[width=10cm]{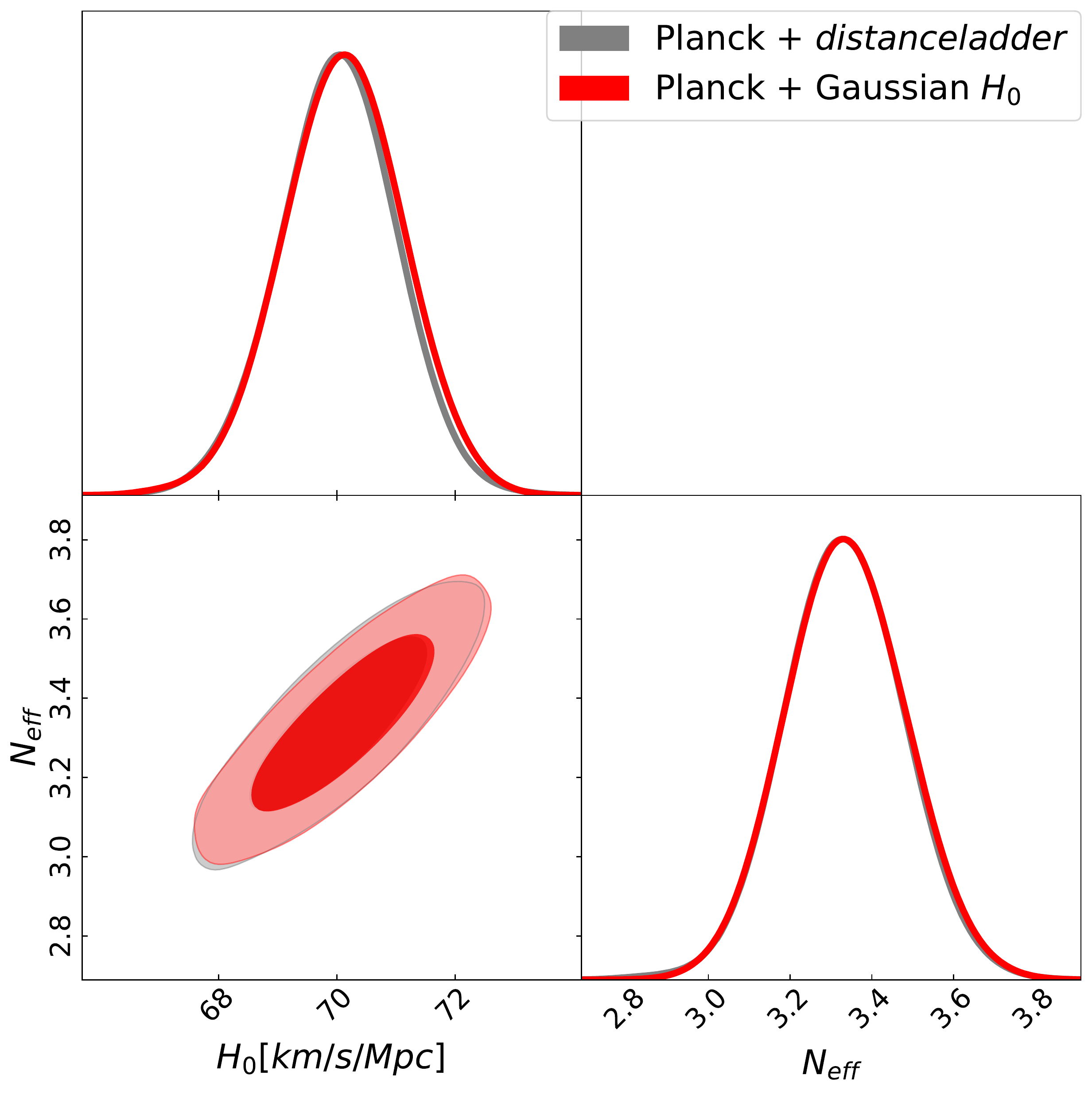}
\centering
\caption{Marginalized posterior distributions for $H_0$ and $N_{\rm eff}$ using either our \textit{distanceladder} likelihood (DL) or a Gaussian $H_0$ prior, in combination with Planck temperature and polarization data for both cases \cite{Planck:2018vyg}. The strong agreement between the \textit{distanceladder} and the Gaussian prior results supports that early-time changes to cosmology only affect distance ladder measurements through $H_0$.}
\label{fig:neffvs}
\end{figure}

To perform this test, we run MCMC anlyses with the code \texttt{MontePython} \cite{Brinckmann:2018cvx}, letting the 6 standard $\Lambda$CDM parameters and $N_{\rm eff}$ vary (as well as necessary nuisance parameters) in the fit. 
As shown in figure \ref{fig:neffvs}, we find that our \textit{distanceladder} likelihood reproduces the results obtained from a Gaussian prior on $H_0$ when performing a joint analysis with CMB data for this one-parameter extension of $\Lambda$CDM. 
In detail, when combined with Planck high-$\ell$ TTTEE, low-$\ell$ EE, and low-$\ell$ TT data \cite{Planck:2018vyg}, our \textit{distanceladder} likelihood finds $N_{\rm{eff}} = 3.340 \pm 0.148$ ($H_0 = 69.96 \pm 1.01$ km/s/Mpc), which is to be compared with $N_{\rm{eff}} = 3.344 \pm 0.149$ ($H_0 = 69.95 \pm 1.02$ km/s/Mpc) when using the Gaussian prior on the Hubble constant. 
The agreement between the \textit{distanceladder} and the Gaussian $H_0$ prior, in this case, demonstrates that the latter is sufficient when analyzing models modifying the cosmological expansion history at early times. 
However, as we will see in the next section, this is generally not the case for models changing the Hubble expansion history at late times. 

\section{Models Impacting Late-time Cosmology: Issues with Gaussian $H_0$ Prior} \label{sec:LDE}
Our \textit{distanceladder} likelihood package allows one to concentrate on the key issue behind the Hubble-Lema\^{i}tre Tension: the distances to SNe Ia in the Hubble flow. 
Refocusing the discussion onto distance disagreements immediately begins to narrow the field of possible solutions to the Hubble-Lema\^{i}tre tension, specifically those which propose changes to late-time cosmology.
Such models may seem at the surface to resolve the tension through their accommodation of larger $H_0$ values compared to $\Lambda$CDM. 
However, they generally do so at the cost of worsening the tension with the actual distance measurements, indicating that such models do not fundamentally address the root cause of the discrepancy and are therefore only of modest interest. 

As an example of how a Gaussian prior on $H_0$ standing in for the entirety of the local distance ladder can be misleading, we consider below a late-time dark energy (LDE) model in which a sudden transition in the abundance of dark energy increases the value of the Hubble rate close to $z=0$ \cite{Benevento:2020fev}. Using our \textit{distanceladder} likelihood package, we show below that such LDE models with a high $H_0$ value poorly fits the measured SNe Ia distances. While we focus here on this simple phenomenological LDE model, we note that other proposed models such as chameleon dark energy \cite{Karwal:2021vpk,Cai:2021wgv} and phenomenological emergent dark energy (see e.g.~refs.~\cite{Li:2019yem,Raveri:2018wln}) will also suffer from this problem.

\subsection{Test Case: A Sudden Dark Energy Transition at Late times}

Prior work has demonstrated that introducing a dynamic component of dark energy, particularly at redshifts $z \lesssim 0.02$, successfully increases the value of $H_0$ locally while also preserving the CMB temperature and polarization spectra \cite{Mortonson:2009qq,Zhao:2017cud}. These solutions generally modify the Friedmann equation to allow for a redshift dependent dark energy density component to be included, such as
\begin{equation}
\begin{aligned}
    H^2(z) = & H^2_0[\Omega_{\rm{r}}(1+z)^4 + \Omega_{\rm{m}}(1+z)^3 + \Omega_{\rm{DE}}h(z)+ \\
             & \Omega_k(1+z)^2],
\end{aligned}
\end{equation}
\begin{equation}
    h(z) = \text{exp} \left[ 3 \int_{0}^{\rm{ln}(1+z)} \text{dln}(1+z')(1+w_{\rm{DE}}(z'))\right],
\end{equation}
\begin{equation}
    w_{\rm{DE}} \equiv \frac{p_{\rm{DE}}}{\rho_{\rm{DE}}}.
\end{equation}
where $\Omega_{\rm{r}}$, $\Omega_{\rm{m}}$, and $\Omega_{\rm{DE}}$ are the ratio of the current energy density in radiation, matter, and dark energy to the critical density of the Universe.
$\Omega_{\rm{k}}$ is the underlying curvature density parameter, which we set to 0 here to enforce flatness, $p_{\rm{DE}}$ is the dark energy pressure, and $\rho_{\rm{DE}}$ is the dark energy density.
A subclass of late-time dark energy models which use the above set of equations also allow the phantom regime $w_{\rm{DE}} < -1$. We remind the reader that the limit of $w_{\rm{DE}} = -1$ represents the standard cosmological constant.

\subsubsection{Model Setup}
To illustrate how changes to late-time cosmology in the presence of a Gaussian $H_0$ prior can lead to misleading results, we focus here on the simple phenomenological model presented in ref.~\cite{Benevento:2020fev} in which the dark energy density acquires a time dependence according to
\begin{equation}
    \rho_{\rm{DE}}(z)=[1+f(z)]\Tilde{\rho}_{\rm{DE}},
\end{equation}
where
\begin{equation}
    f(z) = \frac{2\delta}{\Tilde{\Omega}_{\rm{DE}}}\frac{S(z)}{S(0)},
\end{equation}
\begin{equation}
    S(z) = \frac{1}{2}\left[1-\text{tanh}\left(\frac{z-z_{\rm{t}}}{\Delta z}\right)\right],
\end{equation}
where $\Tilde{\rho}_{\rm{DE}}$ and $\Tilde{\Omega}_{\rm{DE}}$ are the dark energy density and density parameter in a standard $\Lambda$CDM model. In this model, the dark energy density suddenly increases at redshift $z_{\rm{t}}$ as compared to its $\Lambda$CDM value by a fractional factor of $2\delta$, with $\Delta z$ parameterizing the duration of the transition. As a result, this model boosts the $z=0$ Hubble rate according to
\begin{equation}
    H_0^2 = (1+2\delta)\Tilde{H}_0^2,
\end{equation}
where the tilde indicates the original $\Lambda$CDM value.

The critical feature of this model is that when $z \gg z_t$, it is indistinguishable from the $\Lambda$CDM model.
In order to achieve a positive value of $\delta$, the equation of state must be allowed to enter the phantom regime of $w < -1$, which can make a physical implementation of such a phenomenological model difficult.
With its sudden late-time transition, this scenario nevertheless achieves the desired result of `hiding' the reduced distance to the CMB last-scattering surface that usually results from increasing $H_0$. Indeed, using CMB and BAO data in conjunction with a Gaussian $H_0$ prior (based on the results of ref.~\cite{Riess:2019cxk}), ref.~\cite{Benevento:2020fev} showed that such model could achieve $H_0 = 72.5 \pm 1.85$ km/s/Mpc, hence apparently alleviating the tension.

\subsubsection{Taking into account the distance measurements: Results}
\begin{figure}[t] 
\includegraphics[width=11cm]{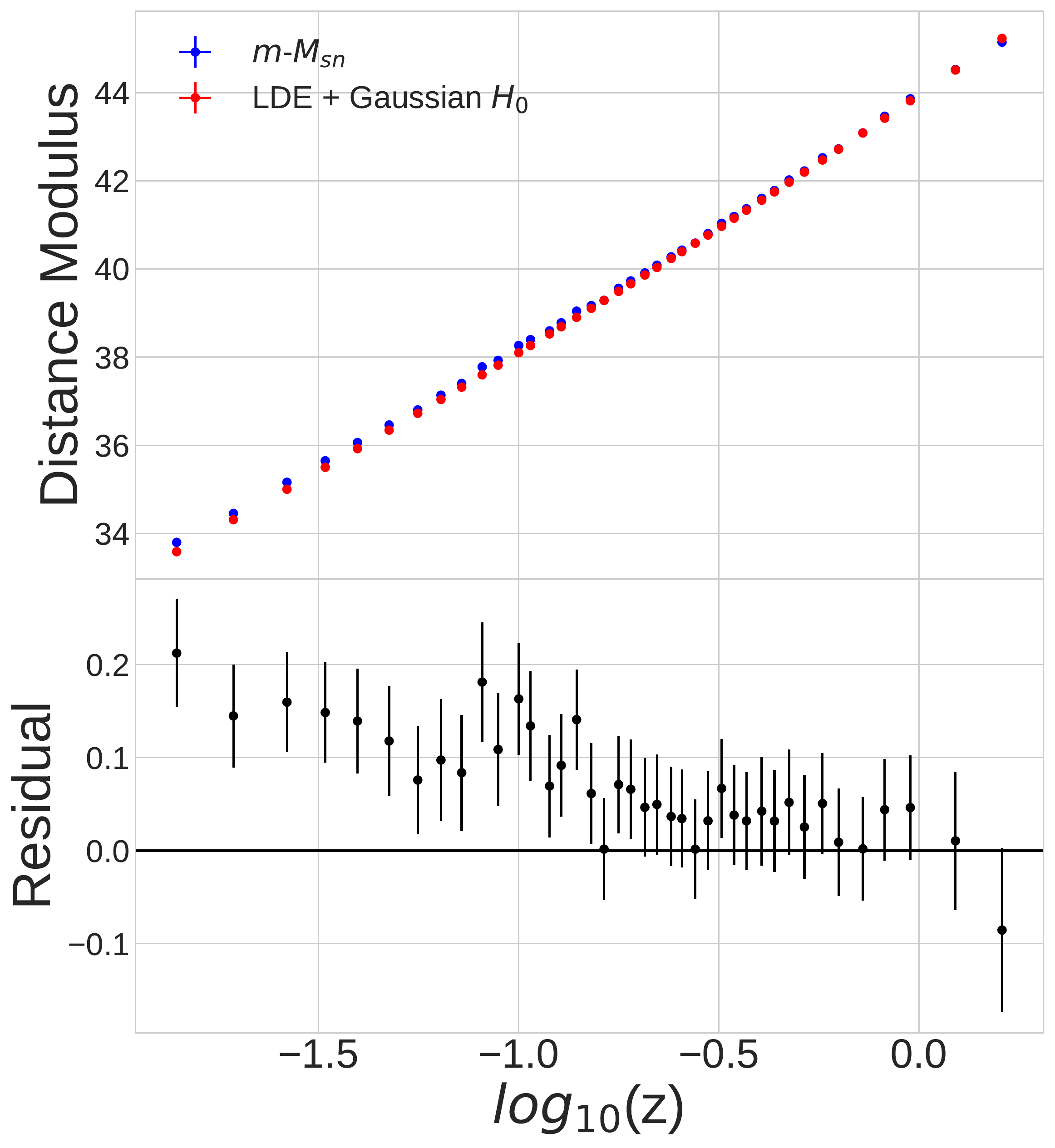}
\centering
\caption{Comparison of the Cepheid-calibrated SNe Ia distance moduli from the Pantheon sample (blue) to the distance moduli predicted by a LDE model fitted with Planck CMB data, BAO data, and a Gaussian prior on $H_0$ (red). The residual difference between the measurements and the LDE predictions are shown in the bottom panel. The error bars on the blue and red points are smaller than the points themselves and are not easily seen. A systematic disagreement is seen between the Cepheid-calibrated distances and the LDE prediction with a Gaussian $H_0$ prior, especially at lower redshifts.}
\label{fig:LDEfail}
\end{figure}

From the analysis quoted above, the LDE model considered here may seem like a promising way to resolve the Hubble-Lema\^{i}tre Tension. However, such models are much less promising once the local distance ladder is properly accounted for \cite{Benevento:2020fev}. As explained above, LDE-based solutions can increase the value of $H_0$ while causing a negligible fractional change in distance to the CMB last-scattering surface and thus to $\theta_{\star}$. However, the fractional change in distance to low-redshift SNe Ia from such late-time transition is, in general, not negligible. This can result in a poor fit to the actual distances, with the lowest redshift objects being the most affected.  
Therefore, LDE models, in general, are ideal examples of being too focused on modifying the value of $H_0$ itself while entirely missing the actual source of the tension (i.e.~the distances). 

\begin{figure}[t!] 
\includegraphics[width=10cm]{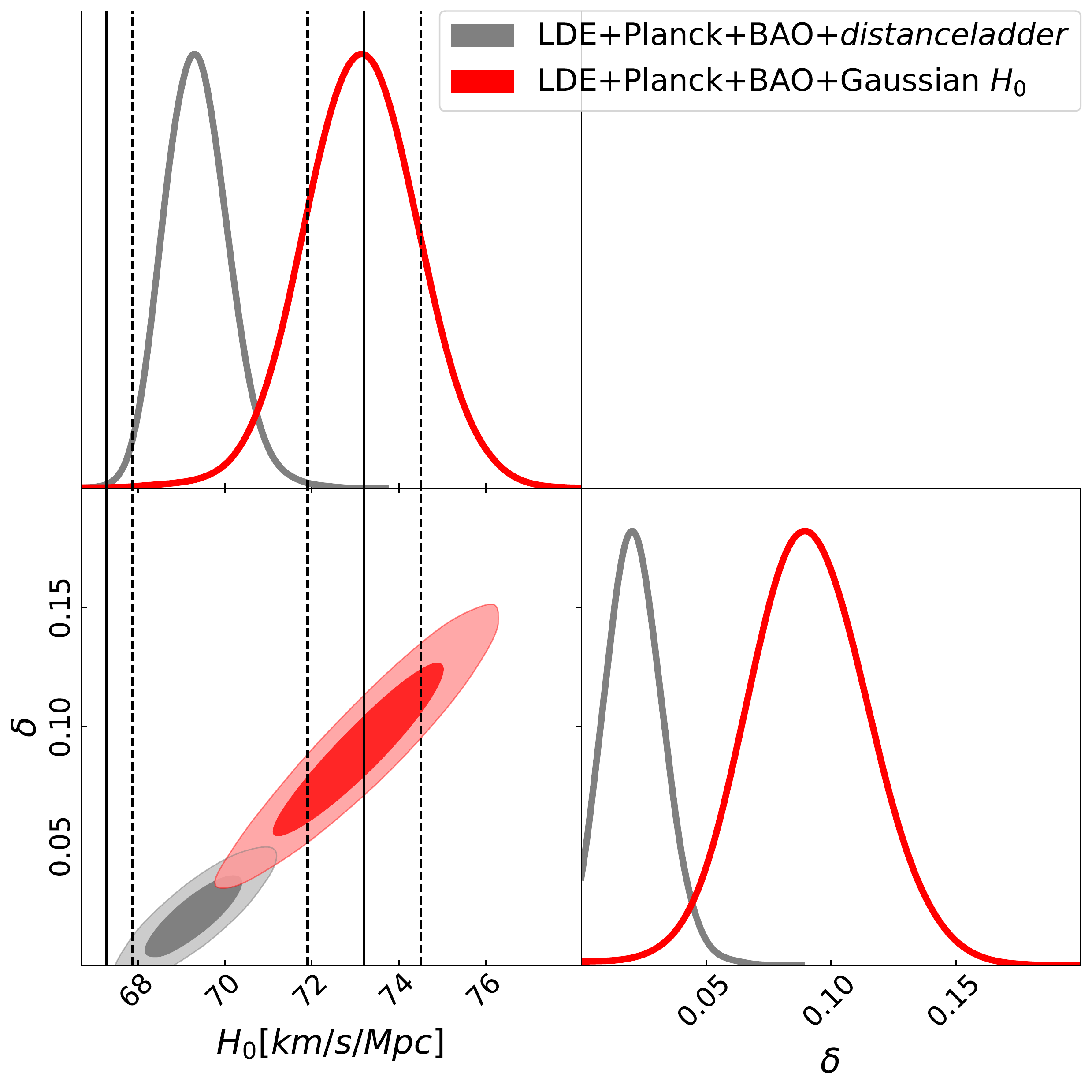}
\centering
\caption{Posterior distributions involving a late-time dark energy model using the traditional Gaussian prior on $H_0$ indicated by red, and $\textit{distanceladder}$ likelihood package in grey. The solid black lines indicate the mean value from ref. \cite{Planck:2018vyg} (left) and ref. \cite{Riess:2020fzl} (right) with the dashed lines indicating the 1$\sigma$ differences.}
\label{fig:Gaussvs}
\end{figure}

The red line in figure \ref{fig:LDEvs} shows the distribution of intercepts values $\beta$ (using the Cepheid calibration) for a LDE model fitted to Planck data and a Gaussian prior on $H_0$ from the R21 measurement.
Recall that for no tension to exist, the intercept should be identically 0.
In this case, the intercept value of the LDE model is $\beta = 0.073\pm0.04$, which is consistent with zero at less than two sigmas. 
While this seems to alleviate the tension between CMB and the Cepheid calibration, this solution is much less promising when looking at the individual distances to nearby supernovae.
This is shown in figure \ref{fig:LDEfail} where we observe a clear offset at low redshifts ($z\lesssim0.1$) between the actual distances to SNe Ia calibrated with Cepheids (blue points), and the corresponding distances predicted in the LDE model (red points) fitted with a Gaussian prior on $H_0$. 
The residual plot in the lower panel of figure \ref{fig:LDEfail} describes the difference in distance modulus required to bring the LDE-predicted SNe Ia distances into agreement with their measurements.
We can further quantify this low-redshift distance tension by fitting the relationship between the predicted distance moduli $\mu$ and the measured $m-M_{\rm sn}$ values with a linear regression with a free slope. Recall that if this slope differs from unity, it would indicate a breakdown of the standard relationship between absolute luminosity and measured flux in an expanding Universe. Performing such fit, we find a slope of $0.983\pm0.002$, which differs from unity at 8.5 $\sigma$ indicating a clear issue in this late-time dark energy model. The corresponding intercept is $\beta = 0.729\pm0.09$, which differs significantly from zero at several sigmas. Clearly, a late dark energy transition does nothing to address the root cause of the tension and can even make the global fit to cosmological data worse. 
In fact, comparing the LDE model fitted with a Gaussian prior on $H_0$ to the LDE model fitted with the \textit{distanceladder} likelihood results in a $\Delta\chi^2 \sim21$, demonstrating that the former provides a poorer fit to the distances than the latter. While these results are broadly consistent with those presented in ref.~\cite{Benevento:2020fev}, our work clarifies why these models fail to properly resolve the tension: they are unable to fit the actual distance to low-redshift SNe Ia in the Hubble flow, as shown in figure \ref{fig:LDEfail}.

Perhaps even more concerning is that using a Gaussian prior, in this case, could lead to spurious detection of new physics. 
As shown in figure \ref{fig:Gaussvs} (see also table \ref{tab:dlvsgauss}), the LDE model analyzed with Planck and BAO data in addition to a Gaussian $H_0$ prior leads to a significant detection ($>3.5\sigma$) of a non-vanishing $\delta$ value, indicating a strong statistical preference for a late transition in the abundance of dark energy. 
However, when the actual calibrated distance measurements are taken into account as in our \textit{distanceladder} likelihood, the value of $\delta$ becomes entirely consistent with zero, and no significant evidence for a late-time transition in the abundance of dark energy is observed. 
In this case, the failure of the Gaussian $H_0$ prior can be traced back to the breakdown of the standard cosmographic expansion of the luminosity distance to SNe Ia in the presence of a sharp transition in the Hubble rate at late times. 

\begin{table}[t!]
\begin{center}
 \begin{tabular}{||c | c c||} 
 \hline
 LDE & Distance Ladder $H_0$ & Gaussian $H_0$ \\ [0.5ex] 
 \hline\hline
 100$\theta_{\rm{s}}$    & 1.04200$\pm$0.00029 & 1.04170$\pm$0.00029 \\ [1ex]
 \hline
 $\Omega_{\rm{b}}$       & 0.02246$\pm$0.00014 & 0.02247$\pm$0.00014 \\ [1ex]
 \hline
 $\Omega_{\rm{c}}$       & 0.11867$\pm$0.00094 & 0.11890$\pm$0.00097 \\ [1ex]
 \hline
 $\tau$                  & 0.0570$\pm$0.0076   & 0.0596$\pm$0.0073 \\ [1ex]
 \hline
 $\rm{ln}(10^{10}A_{\rm{s}})$ & 3.050$\pm$0.015 & 3.052$\pm$0.015 \\ [1ex] 
 \hline
 $n_{\rm{s}}$            & 0.9694$\pm$0.0038 & 0.9689$\pm$0.0038 \\ [1ex] 
 \hline
 $\delta$                & 0.022$\pm$0.022 & 0.091$\pm$0.048 \\ [1ex] 
 \hline\hline
 $H_0$                   & 69.74$\pm$1.28 & 73.78$\pm$2.15 \\ [1ex] 
 \hline
\end{tabular}
\end{center}
\caption{Table describing the mean and 68\% confidence intervals of a MCMC in which the \textit{distanceladder} likelihood and the standard Gaussian likelihood on $H_0$ are used. Prior limited implies that the value is unconstrained within the prior. $\Delta z$ and $z_{\rm{t}}$ are prior limited in all cases.}
\label{tab:dlvsgauss}
\end{table}

\subsection{Discussion}

The above section demonstrates that a Gaussian prior on $H_0$ can lead to a statistically significant detection of a dark energy transition at very late times.
However, once the actual distances to SNe Ia are properly taken into account, such preference largely disappears. Figure \ref{fig:LDEfail} clearly illustrates why this is happening: the Gaussian prior on $H_0$ leads to distances to low-redshift SNe Ia that are too small compared with their actual measured values. Ultimately, this problem arises because the LDE models break the assumptions (in particular, the cosmographic expansion of the luminosity distance) that were used to infer $H_0$ from carefully calibrated distances to SNe Ia in the Hubble flow \cite{Riess:2016jrr,Riess:2020fzl}. This should serve as a cautionary tale when analyzing models that modify the Universe's expansion history at very late times. In general, we recommend that cosmological models that deviate significantly from $\Lambda$CDM at $z\lesssim1$ should never be analyzed with a Gaussian prior on $H_0$. Instead, the calibrated distances to SNe Ia in the Hubble flow should directly be used in statistical analyses of such models. Our \textit{distanceladder} likelihood package make such analyses straightforward.  

\begin{figure}[t!] 
\includegraphics[width=10cm]{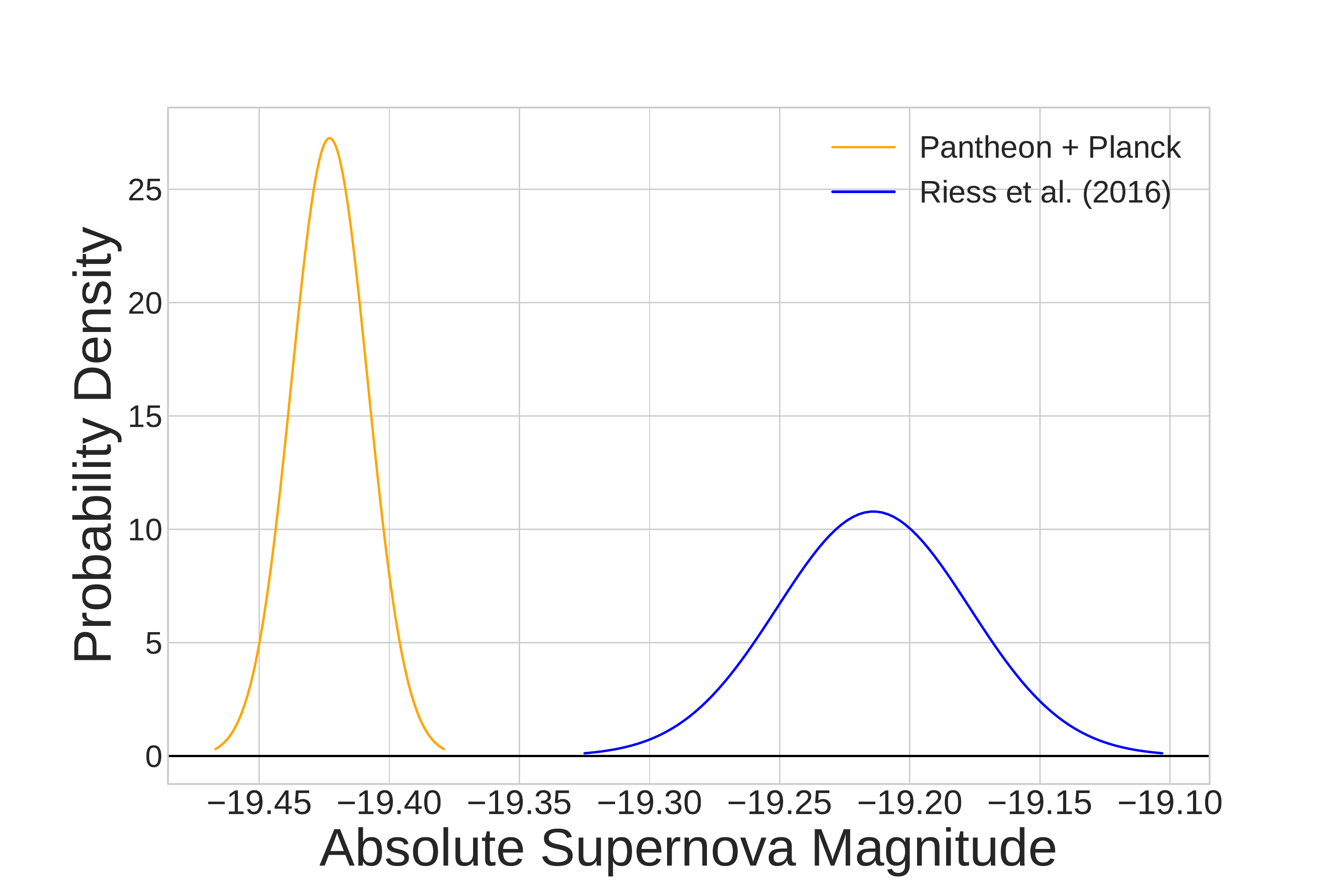}
\centering
\caption{Comparison of the values of $M_{\rm{sn}}$ gotten by calibrating the Pantheon supernova data set with Planck data in $\Lambda$CDM with that obtained from the local distance ladder \cite{Riess:2016jrr}. }
\label{fig:snmag}
\end{figure}

Beyond the possible breakdown of the cosmographic expansion, another issue that has plagued many cosmological analyses using a Gaussian $H_0$ prior in conjunction with CMB data is the resulting ambiguity in the underlying absolute SNe Ia calibration.  
For example, R16 finds a value of $M_{\rm{sn}} = -19.214\pm0.037$.
Imposing the corresponding Gaussian prior on $H_0$ implicitly assumes this $M_{\rm{sn}}$ value, which does not necessarily agree with the value of $M_{\rm{sn}}$ implied by the other likelihoods used.
For instance, the Pantheon data set calibrated with Planck data finds a value of $M_{\rm{sn}} = -19.420\pm0.014$ in $\Lambda$CDM, which is statistically inconsistent with the above value as shown in figure \ref{fig:snmag}. However, since most of the attention is focused on $H_0$ rather than the distances themselves, this underlying inconsistency is often missed in cosmological analyses. Of course, this $\beta\sim-0.2$ mag discrepancy between $M_{\rm{sn}}$ values \emph{is} the root cause of the Hubble-Lema\^itre tension within $\Lambda$CDM. This point was recently emphasized in refs.~\cite{Lemos:2019483,Efstathiou:2020wxn,Camarena:2021jlr,Efstathiou:2021ocp}, and recent analyses of possible solutions to the tension have used $M_{\rm{sn}}$ instead of $H_0$ as their key parameters (see e.g.~ref.~\cite{Schoneberg:2021qvd}). However, as we have seen in figure \ref{fig:LDEvs}, simply making the values of $M_{\rm{sn}}$ between the local and inverse distance ladders compatible (i.e.~$\beta\sim0$) is not necessarily sufficient to ensure a successful cosmological solution: the model must also fit the actual distances to SNe Ia.

\section{Conclusion} \label{conclusion}
An impressive amount of literature has been written on the Hubble-Lema\^{i}tre tension, and it remains a real possibility that it could be the result of some missing pieces in our cosmological model. 
This impressive literature includes cosmological models that modify the cosmological expansion history at $z<1$.
Using a simple Gaussian prior distribution on $H_0$ to represent the entirety of the local distance ladder, some of these solutions can appear to be promising at relieving the tension.
However, in using such a $H_0$ prior, the user implicitly assumes that the local distance ladder directly measures the Hubble expansion rate at $z=0$. In reality, regardless of calibration choice, the local distance ladder actually measures distances to SNe Ia in the Hubble flow in the redshift range $z \gtrsim 0.02$.
Therefore, to properly assess the effectiveness of cosmological models with modified late-time dynamics at relieving the Hubble-Lema\^{i}tre tension, a simple Gaussian prior can not be used to represent the entire distance ladder.
Instead, a likelihood that reconstructs the distance ladder and the distance measurements to the Hubble flow supernova is required for any model which proposes changes to the late-time Universe.

To aid in this, we have developed a likelihood package aptly named \textit{distanceladder} which focuses on fitting the actual distances to SNe Ia, rather than simply fitting for $H_0$.
The \textit{distanceladder} likelihood uses the actual luminosity distance given by the proposed cosmological model and is therefore sensitive to any changes in cosmology presented. This publicly available numerical package allows the user to test the different assumptions made in analyzing the cosmological distance ladder. It is therefore much more flexible and insightful than simply specifying a prior on the absolute supernova magnitude. 
To demonstrate the effectiveness of the likelihood, we have used the example of a late-time transition in the abundance of dark energy to showcase how focusing purely on $H_0$ can lead to the spurious detection of new physics, and a false sense that the  Hubble-Lema\^{i}tre tension has been relieved.
When the distances are correctly accounted for, such late-time dark energy transition are severely constrained in their ability to resolve the Hubble-Lema\^{i}tre tension.
This behaviour is not limited to only this late-time dark energy model but to any model which dramatically changes the late-time expansion history of the Universe as compared to $\Lambda$CDM. 
In general, the success of a cosmological model should be assessed by how well it can fit the actual distances to Hubble-flow objects (including supernovae, strong gravitational lenses, and gravitational waves standard sirens, among others), rather than whether it can reproduce a specific value of $H_0$. 
We encourage the observational cosmology community to make their actual distance measurements broadly available instead of emphasizing the derived Hubble constant values, which contain assumptions that might not be applicable to all cosmological scenarios. 

Within the standard $\Lambda$CDM model, the Hubble-Lema\^{i}tre tension between Cepheid-calibrated SNe Ia distances and Planck CMB data is now at the critical $5\sigma$ threshold \cite{Riess:2021jrx}. Addressing this crucial cosmological problem requires the community to focus on the root cause of the tension -- the distances to low-redshift Hubble-flow objects -- rather than argue about which value of the Hubble constant might be the correct one.

\acknowledgments
This work was supported by the National Science Foundation (NSF) under grant AST-2008696. We would like to thank the UNM Center for Advanced Research Computing, supported in part by the NSF, for providing the research computing resources used in this work. Part of this work was performed at the Aspen Center for Physics, which is supported by NSF grant PHY-1607611.

\begin{appendices}
\section{distanceladder vs. Gaussian prior on $M_{\rm{sn}}$}
\begin{figure}[t!] 
\includegraphics[width=10cm]{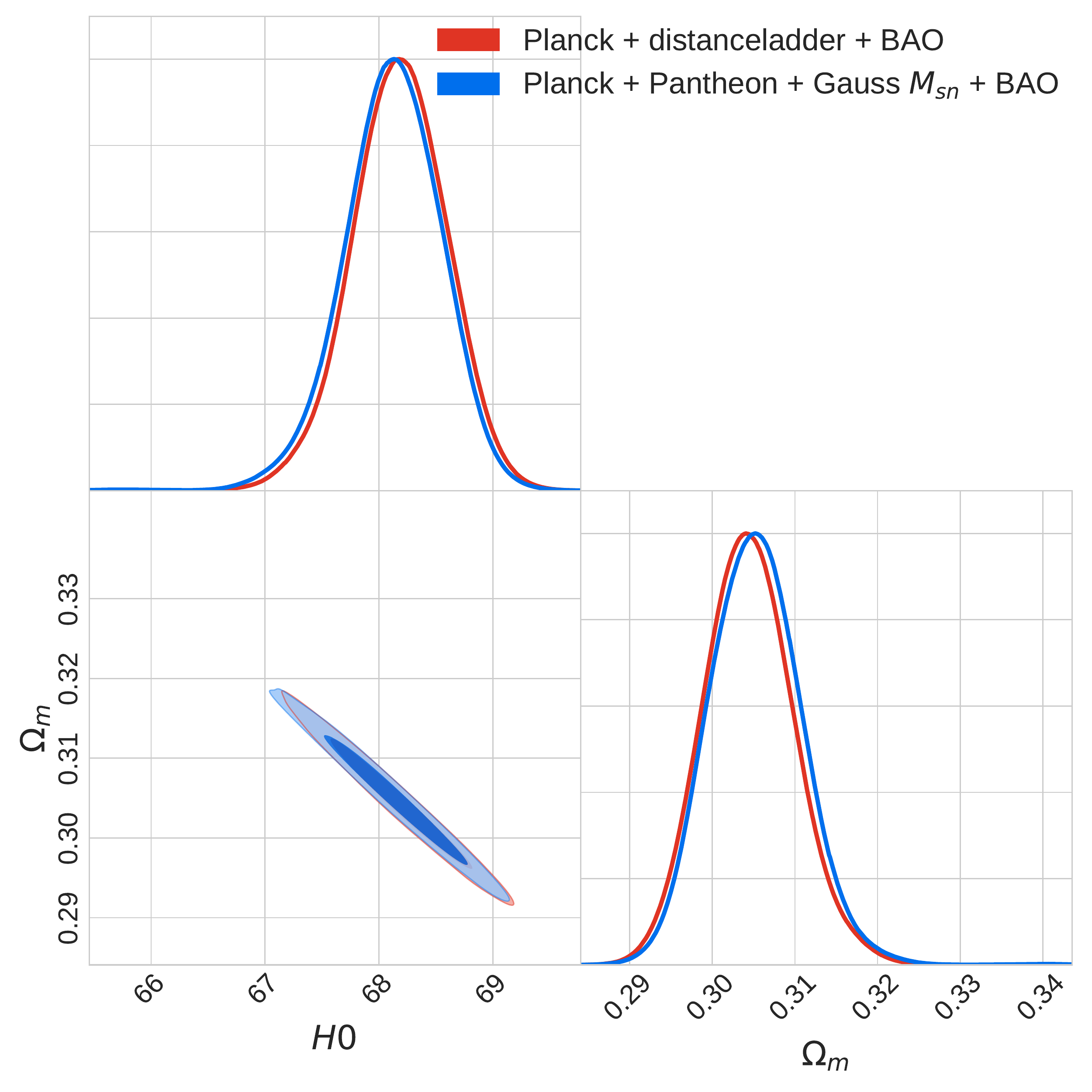}
\centering
\caption{Posterior distributions comparing the results of using the \textit{distanceladder} likelihood and the Pantheon likelihood with a gaussian prior placed on $M_{\rm{sn}}$ assuming a typical $\Lambda$CDM cosmology and including CMB and BAO data. The Planck data utilized are the TT, EE, TE modes and lensing. Here we see a slight difference between the \textit{distanceladder} and the Pantheon likelihood with gaussian $M_{\rm{sn}}$, due to the different handling of the error on $M_{\rm sn}$.}
\label{fig:compare}
\end{figure}

We investigate the differences between using a Gaussian prior on $M_{\rm{sn}}$ along with the Pantheon likelihood and the \textit{distanceladder} likelihood to represent the local measurement of $H_0$.
Both methodologies utilize the distance modulus to compute the maximum likelihood statistics, but make subtle different assumptions in how the covariance matrix is handled.
The Pantheon likelihood with Gaussian $M_{\rm{sn}}$ utilizes a pregenerated covariance matrix with the statistical error on the diagonal.
The \textit{distanceladder} covariance matrix is comprised of three parts.
First, the square of the statistical error along the diagonal associated with measurements of the apparent magintude.
Second, the square of the jacobian which originates from including the error in redshift given by
\begin{equation}
    \delta J = \left(\frac{1+z}{H(z)} + \frac{d_{\rm{L}}}{1+z}\right)\delta z.
\end{equation}
Third, a dense matrix comprised of the square of the error associated with $M_{\rm{sn}}$.
The final covariance matrix for the \textit{distanceladder} is therefore given by
\begin{equation}
\begin{vmatrix}
\delta m_{b,1}^2+\delta J_1^2+\delta M_{\rm{sn}}^2&\delta M_{\rm{sn}}^2&\hdots&\delta M_{\rm{sn}}^2\\
\delta M_{\rm{sn}}^2&\delta m_{b,2}^2+\delta J_2^2+\delta M_{\rm{sn}}^2&\hdots&\delta M_{\rm{sn}}^2\\
\vdots&\vdots&\ddots&\vdots\\
 \delta M_{\rm{sn}}^2& \delta M_{\rm{sn}}^2 &\hdots &\delta m_{b,i}^2+\delta J_i^2+\delta M_{\rm{sn}}^2\\
\end{vmatrix}.
\end{equation}
The results comparing the two methodologies are given in figure \ref{fig:compare}. 
We see a small difference between the two, primarily originating from the handling of the $\delta M_{\rm{sn}}$ error within \textit{distanceladder}. 

\end{appendices}

\bibliographystyle{JHEP.bst}
\bibliography{h0likelihood}

\end{document}